\documentclass{article}

\usepackage{arxiv}
\usepackage[font={small,it}]{caption}
\usepackage[utf8]{inputenc} 
\usepackage[T1]{fontenc}    
\usepackage{hyperref}       
\usepackage{url}            
\usepackage{booktabs}       
\usepackage{amsfonts}       
\usepackage{nicefrac}       
\usepackage{microtype}      
\usepackage{lipsum}
\usepackage{graphicx}
\usepackage{amsmath}
\usepackage[dvipsnames]{xcolor}
\graphicspath{ {./figures/} }
\title{ML-LBM: Machine Learning Aided Flow Simulation in Porous Media}

\author{
  Ying Da~Wang \\
  School of Minerals and Energy Resources Engineering\\
  University of New South Wales\\
  Sydney, NSW, 2052 \\
  \texttt{yingda.wang@unsw.edu.au} \\
  \And
  Traiwit ~Chung \\
  School of Minerals and Energy Resources Engineering\\
  University of New South Wales\\
  Sydney, NSW, 2052 \\
  \texttt{traiwit.chung@unsw.edu.au} \\
  \And
  Ryan T.~Armstrong \\
  School of Minerals and Energy Resources Engineering\\
  University of New South Wales\\
  Sydney, NSW, 2052 \\
  \texttt{ryan.armstrong@unsw.edu.au} \\
  \And
  Peyman~Mostaghimi \\
  School of Minerals and Energy Resources Engineering\\
  University of New South Wales\\
  Sydney, NSW, 2052 \\
  \texttt{peyman@unsw.edu.au} \\
}

\begin{document}
\maketitle

\begin{abstract}
Simulation of fluid flow within complex geometries of porous media has many applications, from the micro-scale (cell membranes, filters, rocks) to macro-scale (groundwater, hydrocarbon reservoirs, and geothermal) and beyond. Direct simulation of flow in porous media requires significant computational resources to solve within reasonable timeframes. An integrated method combining predictions of fluid flow (fast, limited accuracy) with direct flow simulation (slow, high accuracy) is outlined. In the tortuous flow paths of porous media, Deep Learning techniques based on Convolutional Neural Networks (CNNs) are shown to give an accurate estimate of the steady state velocity fields (in all axes), and by extension, the macro-scale permeability. This estimate can be used as-is, or as initial conditions in direct simulation to reach a fully accurate result in a fraction of the compute time. A Gated U-Net Convolutional Neural Network is trained on a datasets of 2D and 3D porous media generated by correlated fields, with their steady state velocity fields calculated from direct LBM simulation. Sensitivity analysis indicates that network accuracy is dependent on (1) the tortuosity of the domain, (2) the size of convolution filters, (3) the use of distance maps as input, (4) the use of mass conservation loss functions. Permeability estimation from these predicted fields reaches over 90\% accuracy for 80\% of cases. It is further shown that these velocity fields are error prone when used for solute transport simulation. Using the predicted velocity fields as initial conditions is shown to accelerate direct flow simulation to physically true steady state conditions an order of magnitude less compute time. Using Deep Learning predictions (or potentially any other approximation method) to accelerate flow simulation to steady state in complex pore structures shows promise as a technique push the boundaries fluid flow modelling.
\end{abstract}
\keywords{Porous Media, Convolutional Neural Networks, Lattice Boltzmann}

\pagebreak
\section{Introduction}
\label{sec:intro}
The flow of fluid within the void space of porous structures is a physical phenomenon that is pervasive in its occurrence, with applications in catalysis \cite{catalysis}, groundwater hydrogeology, oil and gas extraction, environmental waste management \cite{FENWICK1998121,HILPERT2001243,BLUNT20021069,CULLIGAN2006227,mostaghimi2010quantitative, Mostaghimi2016,blunt2017multiphase}, carbon capture storage \cite{BLUNT2013197}, natural and biological membranes \cite{GRUBER2011488}, and medical applications \cite{khanafer2012role}. These examples highlight the need to accurately capture the physics of fluid flow within porous media.

Flow simulation in porous media can be performed with a number of different methods, typically balancing accuracy with speed. Depending on the level of detail required for the application, the pore space can be decomposed dramatically by Pore Network Models (PNM) \cite{BLUNT20021069,pnm1,rabbani2019hybrid}, reducing the computational cost significantly at the cost of obtaining an abstraction of the true velocity and pressure fields averaged over simplified geometries of the pore space. Assuming the pore network itself is generated in a physically representative manner, this is typically adequate for the purposes of permeability estimation \cite{pnmperm}. Conversely, direct flow simulation by solving the Navier-Stokes equation (NVE) explicitly within the pore space offers the highest level of accuracy with the finest degree of detail, projecting directly into the pore space. This can be performed by Finite Method solutions of the NVE \cite{finite6,finite5,finite4,finite3,finite2,finite1}, or by Lattice Boltzmann Methods (LBM) that also solve the NVE \cite{MCCLURE20141865,lbm1,lbm2,lbm3,lbm4,lbm5,lbm6,lbm7}. These techniques require significant compute time and memory. The computational issues with these methods arise due to them being a) time-dependent, and/or b) non-linear/iterative. LBM is time-dependent, NVE is both time-dependent and non-linear, and the simplified steady state stokes equation \cite{peymanK} is iterative. In this study the term "steady-state" refers to the point in which flow simulation converges such that time-stepping and/or iterations become static. Furthermore, in porous media where steady state solutions are sought after, the convergence rate of direct flow simulation to these steady state conditions is also dependent on the geometric complexity of the porous media. A middle ground exists between the PNM and direct simulation in porous media, called Semi-Analytical Solvers (SAS), which are typically derived in the form of a time-invariant, linear, Laplace Partial Differential Equation \cite{pfvs,DGDD,wang2019multi,torskaya2014grain,shabro2012finite}, solvable by Finite Methods. The analytical part of the method arises from the need to characterise a conductivity within the pore space using some geometric approximations or otherwise, leading to a reduction in flexibility and accuracy in certain geometries vs others. As this study aims to learn and predict the true velocity field within porous media, direct simulation is used as it is the most detailed and accurate method. In this case, LBM is used, though Finite solutions of the NVE are similarly accurate. Similarly, due to the approximate nature of PNMs, SASs, and the techniques explored in this study, there exists the potential to use such approximation techniques to predict the velocity field within porous media as a way to speed up the otherwise demanding computation requirement.

From the simulation of flow in porous media, the velocity fields provide necessary information for the determination of critical parameters and phenomena. Of particular interest in this study are a) the permeability, obtained from averaging the velocity fields, and b) the solute transport profile, a fine-scale value obtained by further simulation on top of the velocity fields. Arguably one of the most critical parameter in porous media is the permeability, which is used in the upscaling of the NVE in the pore-scale (typically in the mm to $\mu$m range), to core-scale (cm) and reservoir-scale (m) flow equations based on Darcy's Law. Permeability describes the average flow potential of a porous media, an intrinsic property of the geometry that dictates the relationship between flow and pressure, proportional to the average velocity within the system. As a bulk property, the determination of permeability is relatively insensitive to minor errors and approximations in the velocity field, evident in the reasonable accuracy achieved by Pore Network Models and Semi-Analytical Solvers. However, the actual fine-scale velocity fields in the principle axes are also important in these applications, for example, to characterise the transport of solute within the fluids \cite{Mostaghimi2016,LIU2018130,LIU201712,LIU2017121}. These fine scale velocity fields can be obtained by the direct simulation of flow within the resolved pore space of porous media.

Deep Learning in the form of Neural Networks are used in a wide range of problems, as a form of predictive modelling, or as a form of data processing. Convolutional Neural Networks (CNNs) \cite{SCHMIDHUBER201585} are common in image based problems such as image-to-image translation (of which this study technically can be classified as) \cite{pix2pix2016,cyclegan}, super resolution \cite{EDSR,SRGANledig}, image classification \cite{resnet}, and semantic segmentation \cite{maskrcnn,ronneberger2015unet}. Face/object detection \cite{facedetection}, autonomous driving \cite{segnet}, speech recognition \cite{speech}, and medical image processing/analysis \cite{ldct1,ldct2,ldct3,umeharaSRCNNmedical,ctmedicalresolution1,circleGAN,segITO,segMed3DLI202075} are examples of tasks that these techniques excel at. Specific to porous media, CNNs have been used for super resolution \cite{wang2019super,wangedsrgan,DRSRD1,DeepRock-SR,peymanTSR}, and binary and multi-mineral segmentation \cite{computers8040072,KARIMPOULI2019142,uresnetseg}.

In the specific task of predicting fluid flow and fluid flow properties, CNNs have been applied to reasonable success. These can be categorised roughly into a) the prediction of bulk properties by Regression, and b) the prediction of fine-scale fields by image-to-image translation. The regression of physical properties and fluid flow properties in porous media has focussed on using machine learning and has found success in the prediction of porous media properties \cite{RN112,RN116,RN115,RN114,RN113,ebadi2013,ALQAHTANI2020106514,naifregression} including flow, permeability, porosity, surface area, and other morphological parameters. These methods have shown acceptable accuracy, competitive with traditional approximation techniques for a fraction of the computational cost. The prediction of bulk properties lends itself reasonably well to approximation, with parameters such as the permeability naturally incurring uncertainty in its measurement \cite{chappell2007comparison}. Prediction of fine-scale fields have been applied to predict velocity fields in a number of different applications, including Magnetohydrodynamics \cite{Van_Oort_2019}, Darcy-scale Reservoir Simulation \cite{wang2020efficient}, steady state flow in simple geometries \cite{autodeskflow} and PNMs \cite{rabbani2019hybrid}, and steady state flow in 3D porous media \cite{poreflownet}. In some cases, the network is trained to simply advance a simulation forward \cite{hennigh2017lat, wang2020efficient}, while in others, the steady state scalar fields are predicted in a single pass of the network \cite{poreflownet,autodeskflow,jin2018prediction}. The accuracy achieved in these examples has been shown to be quite high in cases of simple geometries such as vehicle profiles and single objects with simple geometries, though visual discrepancies in the form of noise and velocity profile errors are common. In the few cases studying porous media, the influence that the geometric complexity of the domain (porous or otherwise) has on the accuracy of the network is underexplored. Furthermore, while accuracies are reported as high in terms of the standard pixel/voxelwise measures, the permeability accuracy, and visual comparison, the true usability of these velocity fields for anything more than a visual approximation remains an open question. In cases where the velocity field is predicted as a scalar field, this is more-so the case \cite{poreflownet}. If the network predicts timesteps, errors in the predictions are likely to build up if not corrected for in some manner \cite{wang2020efficient}, while networks trained to directly predict the steady state configuration from only geometric data are prone to higher errors simply due to the extra complexity of the mapping \cite{poreflownet}. These errors may be acceptable when evaluating bulk properties such as the permeability, but may significantly impact usefulness of these predictions if the fine-scale detail is required. In this sense, predictive methods that fall back on the original direct algorithm in a self correcting manner are most effective in preserving accuracy while improving efficiency \cite{hennigh2017lat,wang2019multi}. In the realm of fluid flow in any media of any geometry, solute transport phenomena is one of the most critical simulations that is influenced by the accuracy of the underlying velocity field.

An important aspect to emphasise in this study is the use of feature maps as a part of the input to the network, which is known as Feature Engineering (using encoded features maps as input). In the case of velocity field estimation, feature engineering has shown good results when the trained network is utilised in a patch-based method of prediction \cite{poreflownet} despite each patch having no explicit data regarding the geometry of other regions of the domain. Also, the simple existence of multiple possible flow paths is difficult for a CNN to handle. In the context of velocity field estimation, this local-to-global problem when applying patch-based methods by its nature requires some form of information, in this case, the time-of-flight (TOF) \cite{tof}, which is analogous to the local tortuosity of the geometry \cite{tortuosity}. Without this information, networks trained to predict velocity fields must always be utilised in their global format, which while facilitated by designing the CNN to be fully convolutional (able to be used on any size), can incur higher than acceptable computational cost during prediction. On the other hand, it should be mentioned that calculation of the TOF using the Efficient Fast Marching method scales less favourably to the calculation of the SAS flow approximation or the calculation of Laplace Tortuosity (which utilise the exact same discrete mathematics). Both of these Laplace methods scale O[N] linearly \cite{pfvs,tortuosity} compared to TOF which scales O[NlogN] \cite{tof}. In this case, the SAS Laplace Flow approximation would be the best option as it provides a close, fine-scale approximation to the velocity fields \cite{shabro2012finite,pfvs}. The problem therein lies, if SAS approximations are already quite good, the CNN plays an increasingly lesser role in translating input to output. In the most effective layout, the SAS approximation being transformed to the true LBM approximation is likely the best approach from a Feature Engineering perspective. Since this study also introduces an acceleration procedure for correcting errors in approximated velocity fields, it can also be argued that an SAS approximation can be directly corrected by the acceleration procedure used in this study. 

This study focuses on determining the limitations in the accuracy achievable by end-to-end CNN based image-to-image translation networks on the problem of predicting velocity fields in all principal axes. Specifically, the permeability accuracy and the solute transport accuracy are examined, as one is insensitive to fine-scale fluctuations while the other is highly susceptible. These errors that occur due to the minimisation problem are then corrected for by an accelerated direct simulation using these approximations as input. Porous media is generated in 2D and 3D using correlated fields, and the steady state velocity fields are obtained by Multiple Relaxation Time (MRT-LBM) simulations \cite{MCCLURE20141865}. The networks are trained first in an end-to-end format, with only the binary geometry transforming to the predicted velocity field. This is followed by modifications to the input as a euclidean distance map, the addition of a custom mass conservation function, the removal of biases, and alterations to the aspect ratio of the network (by inversely varying the convolutional filters and kernels). The network achieves an accurate estimate of the steady state velocity fields, measured by permeability and L2 error. Network accuracy is shown to be dependent primarily on the tortuosity (geometric complexity) of the domain. Permeability estimation from these predicted fields reaches over 90\% accuracy for 80\% of cases, but fine-scale velocity fields are error prone when used for solute transport simulation. Using the predicted velocity fields as initial conditions is shown to accelerate direct flow simulation to steady state conditions with an order of magnitude less compute time. Using Deep Learning to accelerate flow simulation to steady state in complex pore structures shows promise as a technique push the boundaries fluid flow modelling.

\section{Methods and Materials}
\label{sec:methods}
\subsection{Datasets}
\label{sec:datasets}

\begin{figure}[htp!]
  \centering
    \includegraphics[width=\textwidth]{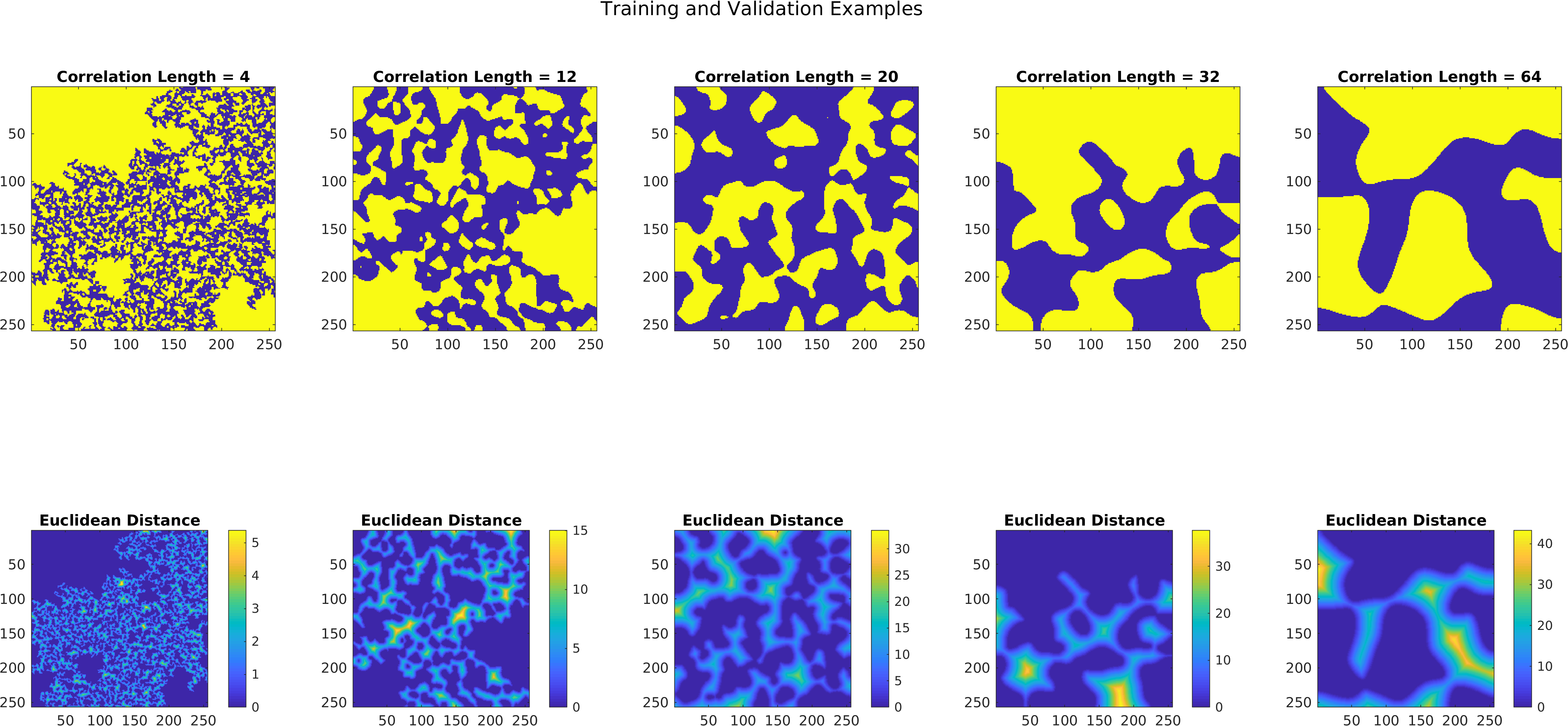}
    \caption{Example 2D images of porous media generated from correlated fields with varying correlation length, and the Euclidean Distance Transform shown below.}
    \label{fig:exampleFields}
\end{figure}

To genereralise the findings in this study, porous media is generated in a manner resulting in a uniform distribution of geometric complexities. 10,000 2D correlated fields are generated stochastically, which emulates the internal structure of porous media \cite{LIU2017121}. These correlated fields are generated in a similar fashion to other such synthetically generated procedures. The algorithm follows a process that involves (1) generating a random field of uniformly distributed numbers, (2) applying a Gaussian blurring kernel to the field with a kernel size (or correlation length) chosen in relation to the desired pore channel size, (3) transforming the field into a uniform distribution $\in [0,1]$, and (4) selecting a threshold value to segment the image. In the case of this dataset, 256x256 domains are generated with a correlation length between 8 and 64, whereby the segmentation threshold $T$ is chosen as the first value $T_{i}$ that allows the domain to be connected from the prescribed inlet to outlet plus a small value $\epsilon$ to allow reasonably sized minimum throat diameters in the generated domains. In this dataset, this value $\epsilon$ is equal to 0.03. Some examples are shown in Figure \ref{fig:exampleFields}. Overall, the dataset is designed to represent a diverse array of geometries, ranging from simple wide channels to tight, complex flow paths. The permeability distribution of the entire dataset spans 5 orders of magnitude.

In 3D applications, the simulation domain can exceed 1,000\textsuperscript{3} voxels, which is outside the feasible realm of Deep Learning, which requires relatively smaller domain sizes to maintain fast iterations during the training process. Typically, volumes smaller than 100\textsuperscript{3} are used for training 3D networks \cite{wangedsrgan,poreflownet,autodeskflow,wang2020physical}, since GPUs are limited in available memory, and computational cost scales poorly in 3D. For reference, training the Resnet-like network on a batch size of 4-8x64x64x64 will fit inside 8GB of GPU memory \cite{naifregression}, which scales poorly to other GPUs. An Nvidia RTX Titan contains 24GB of memory, while a Quadro RTX 8000 contains 48GB of memory. Doubling the domain size to 128x128x128 for the same network would incur 64GB of memory on a single GPU. Even a single 256\textsuperscript{3} batch image would not fit on a single GPU. To circumvent this, patch based learning can be applied \cite{poreflownet}, or a fully convolutional network can be trained, allowing the network to be used on domains of arbitrary size, or distributed training on GPU clusters can be performed (though the size would then be limited based on a batch size of 1).

\begin{figure}
  \centering
    \includegraphics[width=\textwidth]{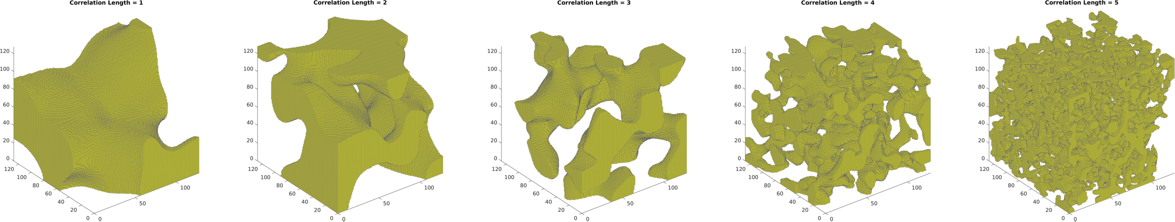}
    \caption{Example 3D images of porous media generated from correlated fields, with varying correlation length.}
    \label{fig:3dsimpcomp}
\end{figure}

In this case, 1,000 3D images of segmented correlated fields of size 128\textsuperscript{3} are also generated for 3D neural network training, using the same methodology as the generation of 2D porous media. Some examples are shown in Figure \ref{fig:3dsimpcomp}

All datasets are generated with both a binary image as input as well as the Euclidean Distance Transform (EDT) of the pore space. Major factors that influence flow fields such as (a) number of connected flow paths, (b) overall tortuosity, and (c) minimum throat radius have significant effects on simulation time and stability. Of these parameters, the minimum throat radius is the predominant factor that causes numerical instability and small time-stepping, resulting in slow computation towards steady state conditions. There are numerous examples within this dataset of tortuous and tight domains where the principal flow path contains a throat with 3-5 voxels in diameter, causing a significant increase in simulation time. One of the objectives of CNN based velocity field estimation is the bypassing of the need for long computation times in domains with restrictive minimum pore throats.

\subsection{Flow Simulation by the Lattice Boltzmann Method}
\label{sec:lbm}

Flow within the pore space is calculated by the Lattice Boltzmann Method (LBM) using a Multi-Relaxation Time (MRT) scheme in D3Q19 quadrature space \cite{wang2019multi}. LBM reformulates the Navier-Stokes Equations (NVE) by numerically estimating the resulting continuum mechanics from underlying kinetic theory. The kinetics of a bulk collection of particles within a control volume is estimated with a vector velocity space $\xi_{q}$ and velocity distributions $f_{q}$. For each velocity space vector $\xi_{q}$, the velocity component in the specified direction is given by $f_{q}$. In 2D, $q=9$ and in 3D, $q=19$. Using these concepts, an equation can be constructed that details the development of fluid transport. In particular, the momentum transport equation at location $\vec{x}_{i}$ over a timestep $\delta t$ takes the form in Eqn \ref{eqn:lbmMomentum} that relies on a collision operation $J$ which recovers the Navier Stokes Equation, and outlined in detail in \cite{MCCLURE20141865}:

\begin{equation}
\label{eqn:lbmMomentum}
f_{q}(\vec{x}_{i}+\vec{\xi_{q}}\delta t, t+\delta t) = f_{q}(\vec{x}_{i}, t) + J(\vec{x}_{i}, t)
\end{equation}

Single phase flow is simulated within the pore space of the segmented test samples until steady state conditions are reached. This is measured by tracking changes to the sample permeability $K$ by Eqn \ref{eqn:perm}:

\begin{equation}
\label{eqn:perm}
K=\frac{\mu \bar{\vec{v}}L}{\nabla P_{x}}
\end{equation}

where $\mu$ is the kinematic viscosity, $\bar{\vec{v}}$ is the mean velocity within the bulk domain, $L$ is the length of the sample in the direction of flow, and $\Delta P$ is the pressure difference between the inlet and outlet. The permeability of any given porous media is a constant value at steady state conditions, when the velocity fields and the pressure fields become time-static. In this case, simulations are run until the change in permeability over 1,000 LBM timesteps $\frac{\delta K}{\delta t}|_{t=1000}$ is less than 1e-5. All samples are simulated with a constant pressure drop between the inlet and outlet, set such that the mean pressure gradient is 1e-5, and wall boundary conditions are imposed along the other sides to avoid geometric inconsistencies associated with periodic boundary conditions. The velocity fields are channel concatenated together, and used as ground truth output during training.

\subsection{Neural Network Architecture}
\label{sec:neuralnetworkarchitecture}

The network formulated for the task of predicting steady state velocity fields takes the form of a U-Net structure \cite{ronneberger2015unet} that contains gated convolutional layers and concatenated activations within residual blocks situated at each node along the U-Net structure. The design is adapted as a combination of features from PixelCNN++ \cite{salimans2017pixelcnn} and from the relatively simple implementation for the purpose of flow field prediction in simple geometries \cite{autodeskflow}. Instead of the usual ResNet-style residual block in each section of the network (which consists of a pair of conv+batchNorm+reLU modules), after the initial convolution at the start of each block, a gated convolution is performed, followed by a skip connection that adds the average-pool into itself. The gated convolution simply consists of 1) convolution with twice the filter number, 2) splitting the output into 2 parts along the filter dimension, 3) applying logistical activation to one of the parts, and 4) multiplying the parts together. 

In each level of the descending portion of the U-net structure, this set of operations is repeated twice, once with a stride of 1, and again with a stride of 2. The outputs from each level are skip connected with the ascending portions of the structure, which consists of transposed convolutions with a stride of 2 followed by one instance of the residual block. The connection is preceded by a fully connected layer that operates only on the filter dimension (a network-in-network), preserving the fully convolutional structure of the network.

Due to the presence of significant regions of sparsity in solid voxels (up to 90\%) where both inputs and outputs equal zero, network biases are disabled in the final configuration. This effect is tested in later sections, and shows marked improvement, as the mapping from the Euclidean Distance Transform (EDT) to the velocity vector space can be mapped multiplicatively.

\begin{figure}[htp!]
  \centering
    \includegraphics[width=\textwidth]{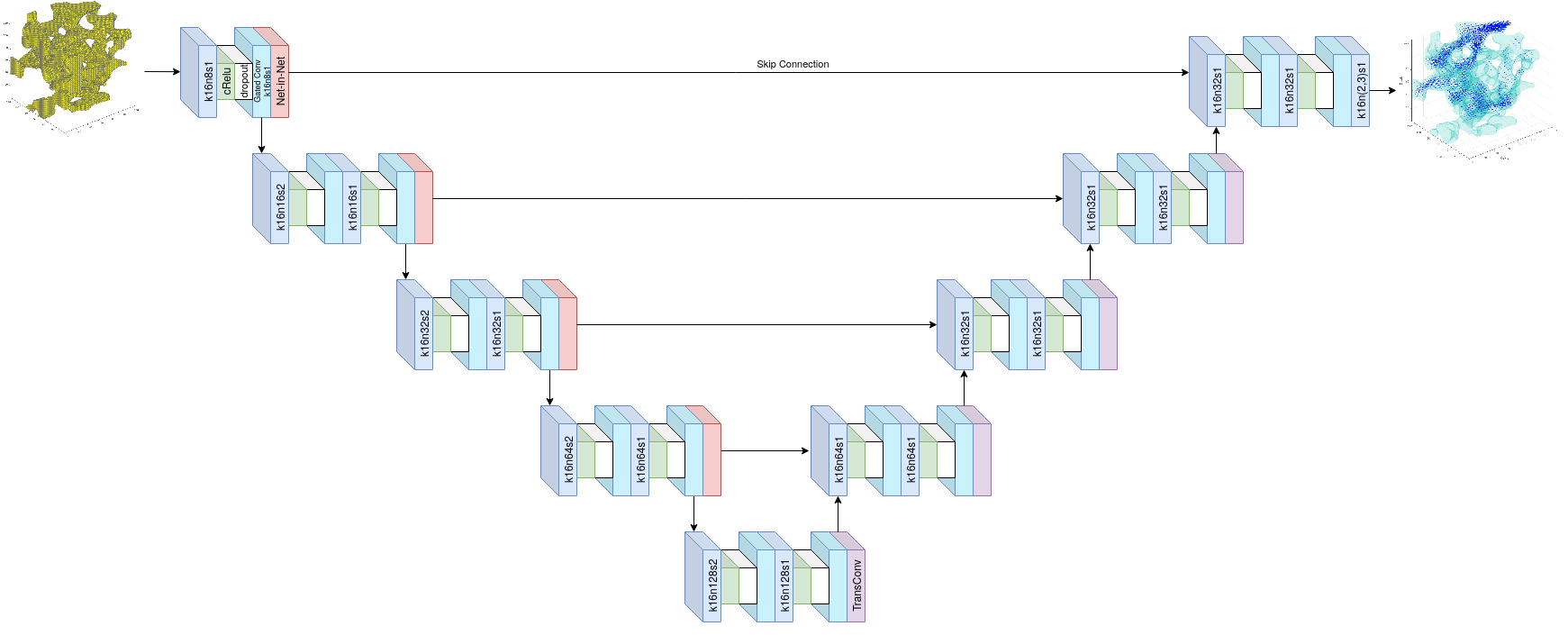}
    \caption{Architecture of the Gated U-net based on PixelCNN++. The U-Net structure is preserved, and gated convolutions are added to each block. A network-in-network layer is also applied before skip connections are applied.}
    \label{fig:uresnet3d}
\end{figure}

In the configuration with a base kernel size of 4 and a base filter size of 32, the total number of trainable parameters is 75M. In 3D, a similar network is used, with 19M parameters to reduce computational cost. The learning rate decayed during training from 1e-5 to 1e-8, and the Adam optimiser is used \cite{AdamKingma}. The network architecture is tested under different configurations of kernel and filter sizes, outlined in subsequent sections. The original binary geometry, or the Euclidean Distance Map is also used as input, and custom physics-based loss functions are used. Other types of feature maps can be added as extra input channels, such as the tortuosity field, time-of-flight, Local Distance Maximum, etc, and though they are mentioned and discussed, are not directly explored in this study.

\subsection{Loss Functions and Accuracy Measures}
\label{sec:accuracy}

In terms of training the network, the primary loss function used is the Mean Squared Error, or the L2 loss between the real field $F_{r}$ and the predicted field $F_{p}$, shown as Eqn \ref{eqn:L2}. Since velocity fields follow a roughly log-normal distribution, sparsely occurring large magnitude velocity values that significantly contribute to mass flux are emphasised during training. Alternatively, one might choose to train such a problem with the Mean Absolute Error, or the L1 loss, as it better accounts for intermediate and small velocity vectors. In this study, the L2 error is chosen as it reduces the largest pixelwise errors faster.

\begin{equation}
\label{eqn:L2}
    L_{2_{Loss}}=\frac{\sum_i(F_{r_{i}}-F_{p_{i}})^2}{\sum_ii}
\end{equation}

In the prediction of velocity fields during steady state flow, a pixelwise loss such as the L2 loss does not explicitly enforce any underlying physics. In order to improve the physical accuracy of the predicted fields, a mass conservation loss is applied. This mass conservation loss $L_{cons}$ is calculated as the mean squared error of all orthogonal flow rate vectors for all velocity components, between the real and predicted field. For example, in 2D, the flow rate $q$ along the X axis consists of the perpendicular component $q_{x}$, calculated by the sum of all velocity vectors in each X slice, summed along the Y axis, and similarly for $q_{y}$. Thus, for a tensorfield of $n$ dimensions, the mass conservation loss is given by Eqn \ref{eqn:Lcons}

\begin{equation}
\label{eqn:Lcons}
    L_{cons}=\sum_n\frac{\sum_i(q_{n_{r}}-q_{n_{p}})^2}{N}
\end{equation}

It should be noted that a mass conservation loss function can alternatively be defined using a pixelwise method, which is similar to a gradient loss function. While the use of a cross-sectional flow-based conservation loss function is observed to be more stable (training can converge with only the $L_{cons}$ active), this is not further explored in this study, and the incremental accuracy improvements are used as-is. 

During testing, as a measurement of mass conservation, the summation L1 version of the mass conservation loss is used, and can be interpreted as the Scaled Total Absolute Flow Error (STAFE), given by Eqn \ref{eqn:STAFE}

\begin{equation}
\label{eqn:STAFE}
    L_{STAFE}=\sum_n\frac{\sum_i|q_{n_{r}}-q_{n_{p}}|}{{\sum_i|q_{N_{r}}}|}
\end{equation}
Where $N$ is the principal direction of flow. Traditionally, porous media flow characteristics at steady state is expressed by the permeability, given by Eqn \ref{eqn:perm}, and the permeability error between the real permeability $K_{r}$ and the predicted permeability $K_{p}$ can be simply calculated as Eqn \ref{eqn:permL1}

\begin{equation}
\label{eqn:permL1}
    L_{Perm}=|1-\frac{K_{p}}{K_{r}}|
\end{equation}

While not a measure of loss or accuracy, the tortuosity of a pore space $a$ can provide valuable quantitative information regarding the geometric complexity of the domain. The tortuosity can be obtained by solving the Poisson Equation within the pore space of the domain \cite{tortuosity} with Dirichlet boundary conditions along the inlet and outlet of 1 and 0 respectively, and calculating the root-mean-square of the divergence field. This process is a simplification of a similar procedure to estimate fluid flow in porous media using the Laplace Equation \cite{pfvs}, and incurs a similar computational cost when solved using Finite Methods with the Algebraic Multi-Grid approach.

\subsection{Accelerating Simulation to Steady State}
\label{sec:acceleratingPerm}

The minimisation problem of Deep Learning is a natural restriction to the ultimate degree of accuracy that can be obtained from predictions using neural networks. As such, it is prudent to couple together such Soft Computing techniques with their rigid counterparts. As such, predicted velocity fields can be used as initial/restart conditions in LBM to accelerate the simulation to its steady state configuration. If the errors associated with pressure fields (which are also predicted during training or by training a separate pressure prediction network) and velocity fields is reasonably small and well distributed, then the propagation of error waves within the initialised simulation domain relaxes quickly to the steady state condition compared to an initialisation of a constant field. 

\section{Results}
\label{sec:results}
The accuracy of various configurations of the network is investigated using 2D datasets, and the best performing configuration is trained on 3D images. The different network configurations aim to determine the relationship between velocity field accuracy and general features of CNNs such as the depth and width of convolutional kernels/filters and the loss functions applied. These overarching parameters apply to some extent to higher dimension problems, as the operations are analogous from 2D to 3D. The accuracy of resulting velocity field predictions is analysed in terms of their overall predictions of permeability as well as their fine-scale accuracy as tested by convection-diffusion transport within the porous media. Acceleration of steady state simulations by using predicted fields as input is also investigated. Training and testing was performed on Nvidia Titan RTX GPUs, and training time for 2D and 3D networks ranged in the 2 day mark.

\subsection{Network Parameter Sensitivity}
\label{sec:kernelsAndFilters}
\begin{figure}[htp!]
  \centering
    \includegraphics[width=\textwidth]{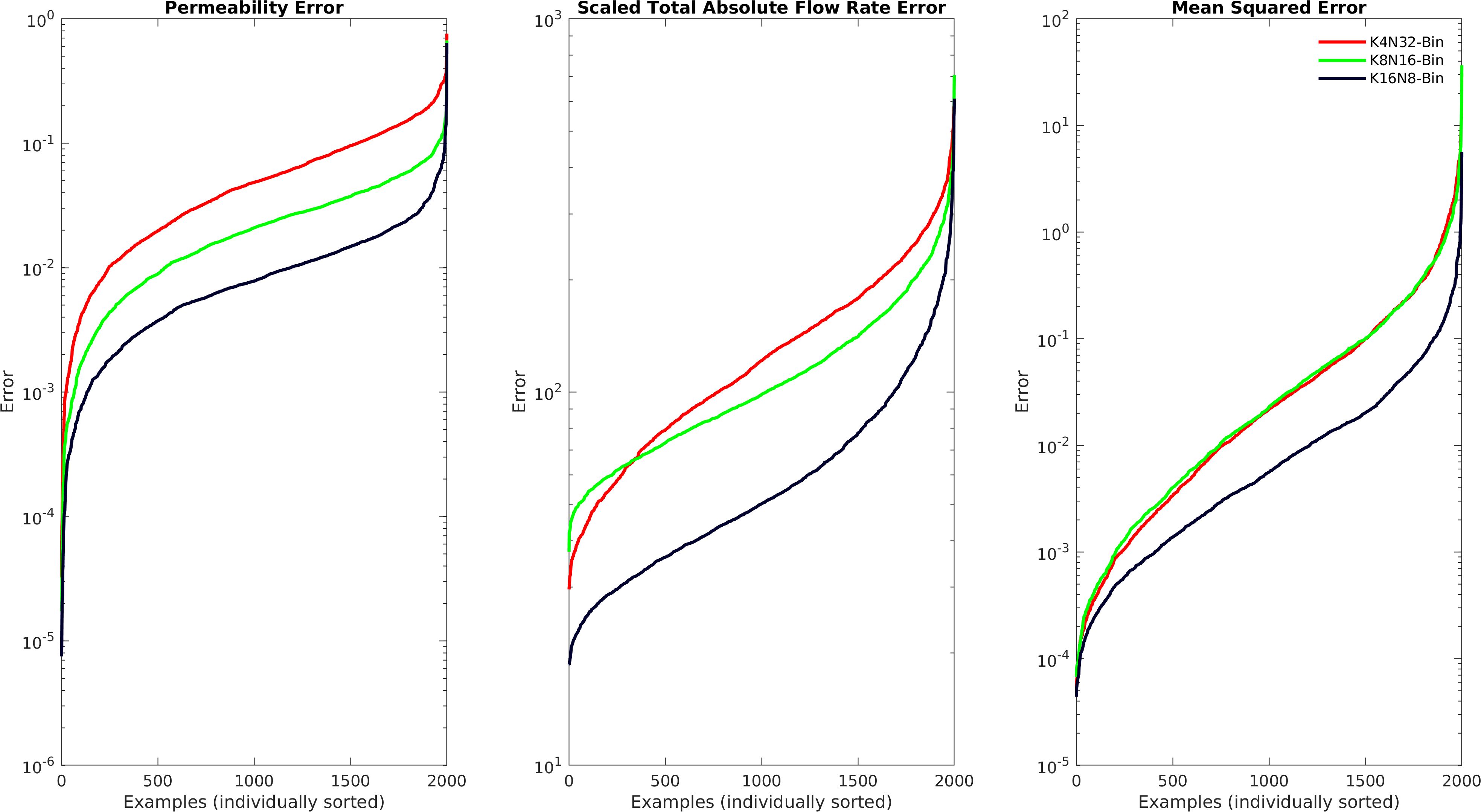}
    \caption{Sorted plots of the accuracy achieved by Gated U-ResNet under different configurations of kernel and filter sizes. The best accuracies achieved over the testing dataset are generated by the K16N8 configuration, reflecting the relatively far influence that wall geometry has on velocity field prediction.}
    \label{fig:configCompare}
\end{figure}

Different kernel sizes and filter numbers are tested using the same architecture to determine the best configuration. The binary image is used as input for this portion of the testing, and the 2D dataset is used. In 2D, varying the kernel size and filter numbers inversely to each other, the network architecture retains the same number of trainable parameters, in this case 76M. Configurations of K4N32 (Kernel Size = 4, Base Number of Filters = 32), K8N16, and K16N8 are tested on 2,000 correlated fields with varying correlation lengths. It should be noted also, that a K32N4 configuration suffered from stability issues during training, likely due to poor numerical scaling from the larger set of kernel weights. The resulting error is calculated using the pixelwise MSE, the permeability error, and the Scaled Total Absolute Flow Error (STAFE). The overall accuracies achieved are shown in Figure \ref{fig:configCompare}, presented as individually sorted graphs for the 3 different configurations.

Results show that the K16N8 configuration performs best based on all accuracy metrics. It appears from this analysis that the accuracy of predicting velocity fields is dependent on the kernel size of the network, which can be translated as feeding a wider field of view of the porous geometry to the network. The kernel size plays a far more important role in velocity field prediction than the filter depth, which is a reflection of the manner in which velocity profiles develop in the pore space as a relation to the surrounding wall geometry. The wider the kernel, the more of this influence-at-a-distance is captured.

A permeability error of less than 10\% is achieved by the K16N8 network over 99\% of the 2,000 correlated fields in the testing dataset. This degree of permeability error is reasonable, and consistent with other types of predictions using regression \cite{RN114}. However, the purpose of predicting velocity fields is to use these fields in analysis that depend on the finer voxel-by-voxel detail afforded by a direct translation of the domain geometry. This is investigated in detail in later sections.

\subsection{Binary Inputs, Mass Conservation, and Bias Removal}
\label{sec:geomvsDistvsMass}

\begin{figure}[htp!]
  \centering
    \includegraphics[width=\textwidth]{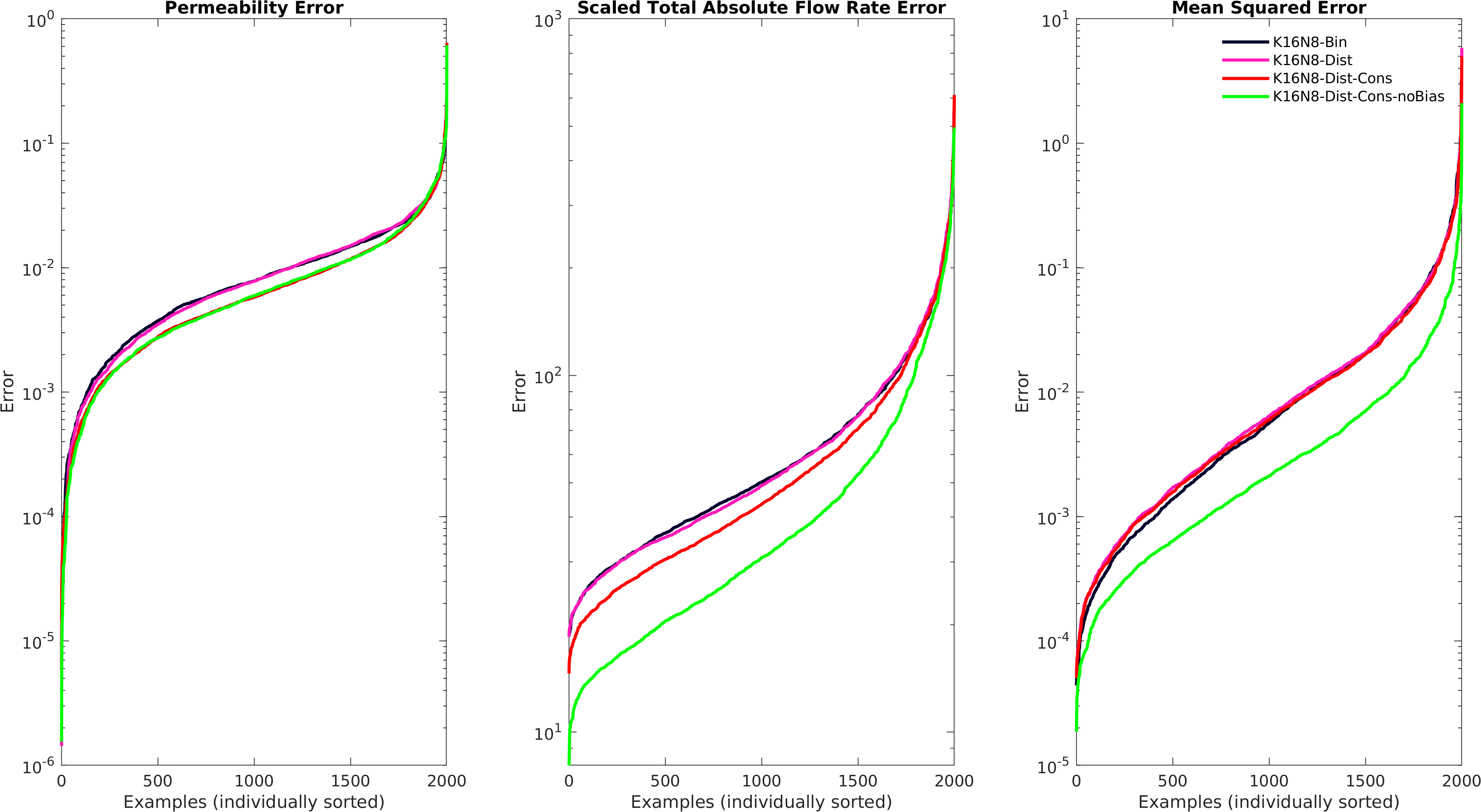}
    \caption{Sorted plots of the accuracy achieved by using (a) EDT as input instead of the binary map, (b) the mass conservation loss, and (c) no biases in layer operations. Errors in the permeability and mass flow errors are lower using the EDT as input, and further lower when using the mass conservation loss function with the MSE. A significant improvement in accuracy is achieved when removing biases from the network, as the sparse porous structure that contains many zeros is operated on exclusively multiplicatively. The improvement in accuracy measures can be attributable to a better match in regions of lower velocity}
    \label{fig:configCompareDist}
\end{figure}

Having established a trend of wider convolutional kernels leading to better overall accuracy, any improvements afforded by using a better encoded input domain is investigated. In this case, the non-linear relationship between the Euclidean Distance Transform (EDT) and the velocity fields within porous media can improve the performance of the network, for minimal extra pre-processing. Using the K16N8 configuration, the training and validation input domains are converted from their binary solid-pore representation to the EDT of the pore space. Similarly, the mass conservation loss function Eqn \ref{eqn:Lcons} is also applied to training, and finally, biases are removed from the network. It is important to explicitly mention at this point, that "Biases" refers to the offset term $b$ within the CNN layers, which would normally perform the linear operation $y=ax+b$. By making these alterations, performance is improved as shown in Figure \ref{fig:configCompareDist}. The use of EDT, mass conservation loss, and removal of biases all contribute towards improving the accuracy of the network when measured by metrics that are sensitive to the fine-scale accuracy, such as the STAFE. These modifications to the network do not appear to significantly affect the permeability, as it tends to average out these errors.

\begin{figure}[htp!]
  \centering
  \begin{minipage}[b]{0.66\textwidth}
    \includegraphics[width=\textwidth]{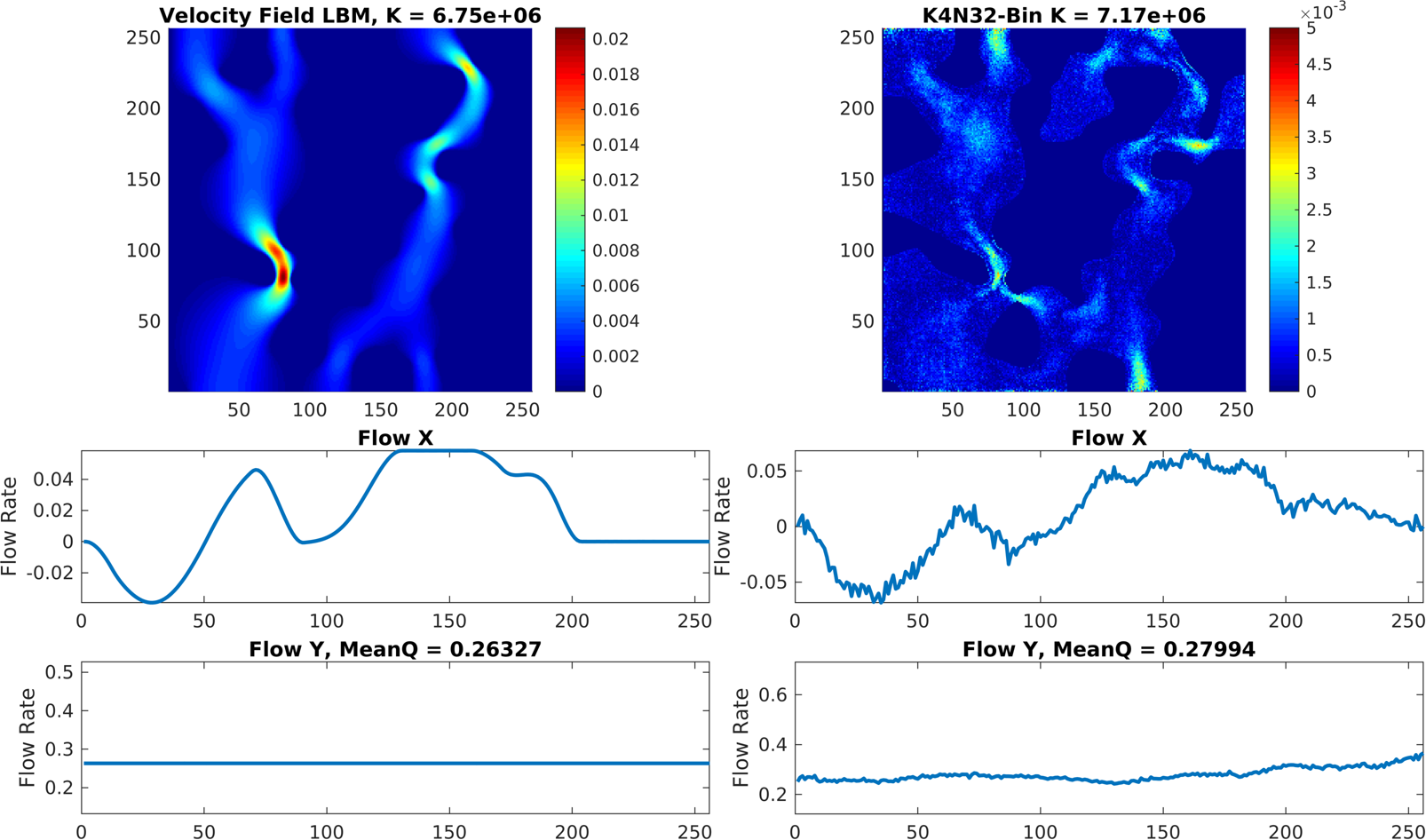}
  \end{minipage}
  \vspace{5mm}
    \begin{minipage}[b]{0.33\textwidth}
    \includegraphics[width=\textwidth]{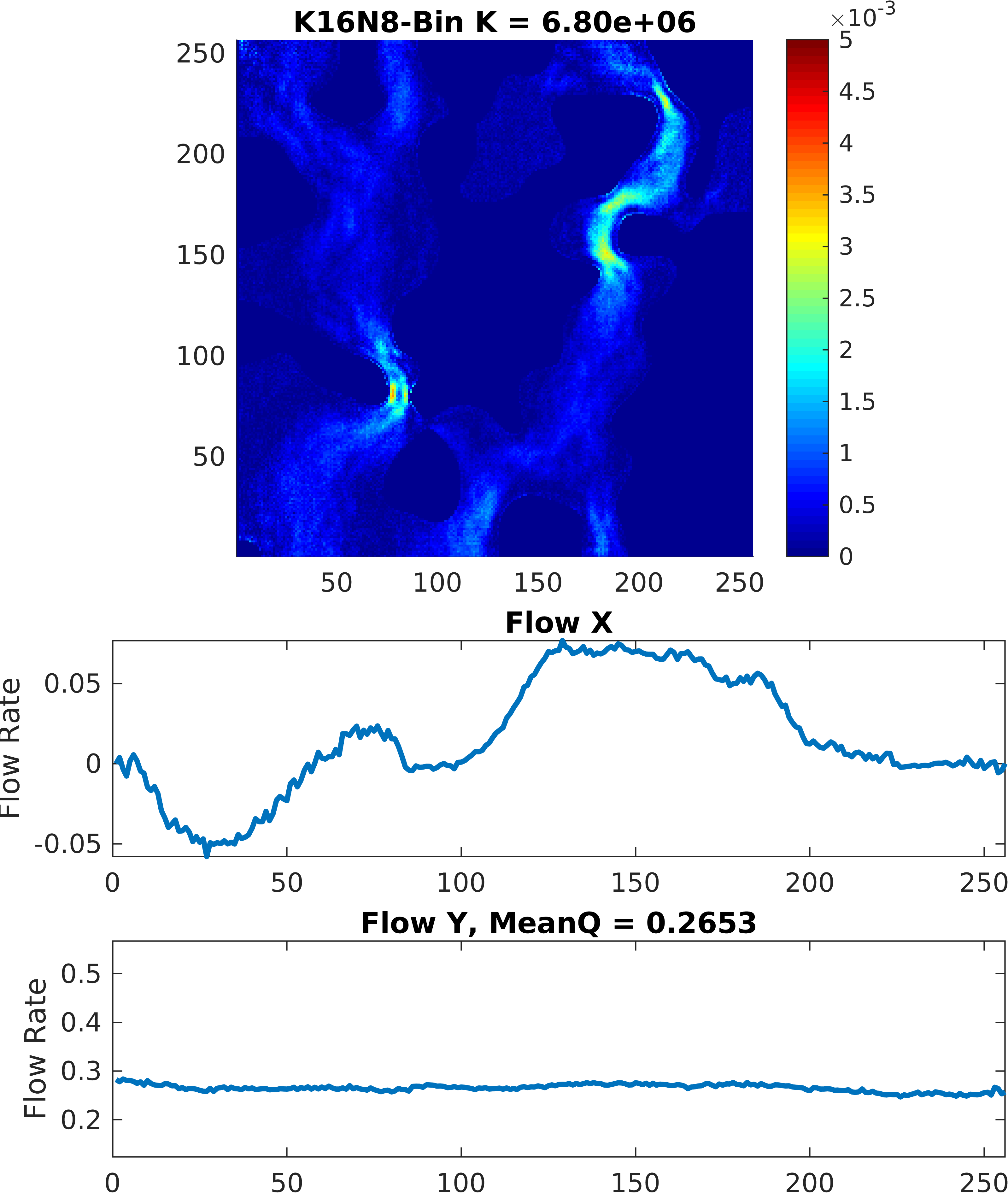}
  \end{minipage}
  \vspace{5mm}
  \begin{minipage}[b]{0.32\textwidth}
    \includegraphics[width=\textwidth]{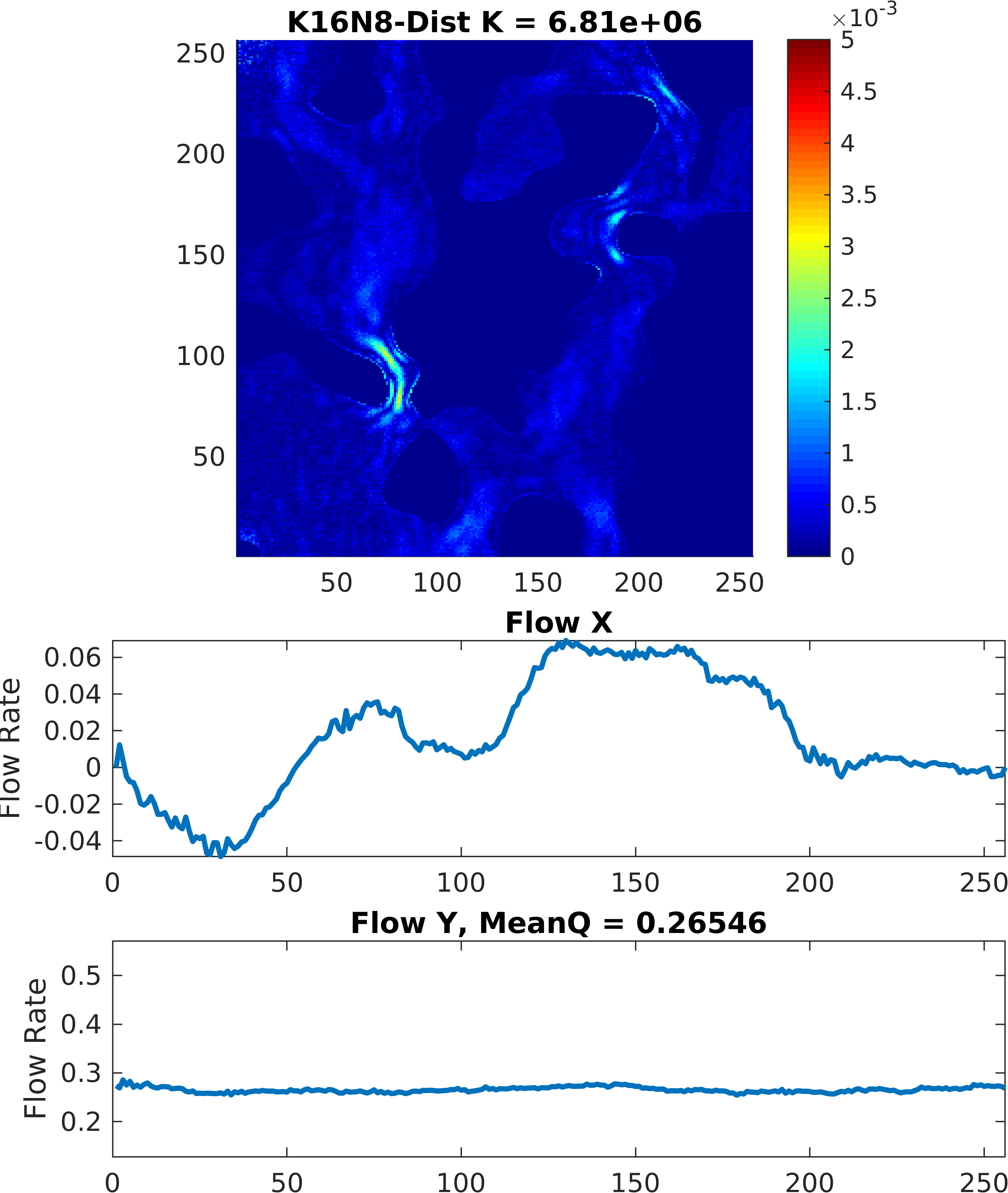}
  \end{minipage}
  \vspace{5mm}
    \begin{minipage}[b]{0.32\textwidth}
    \includegraphics[width=\textwidth]{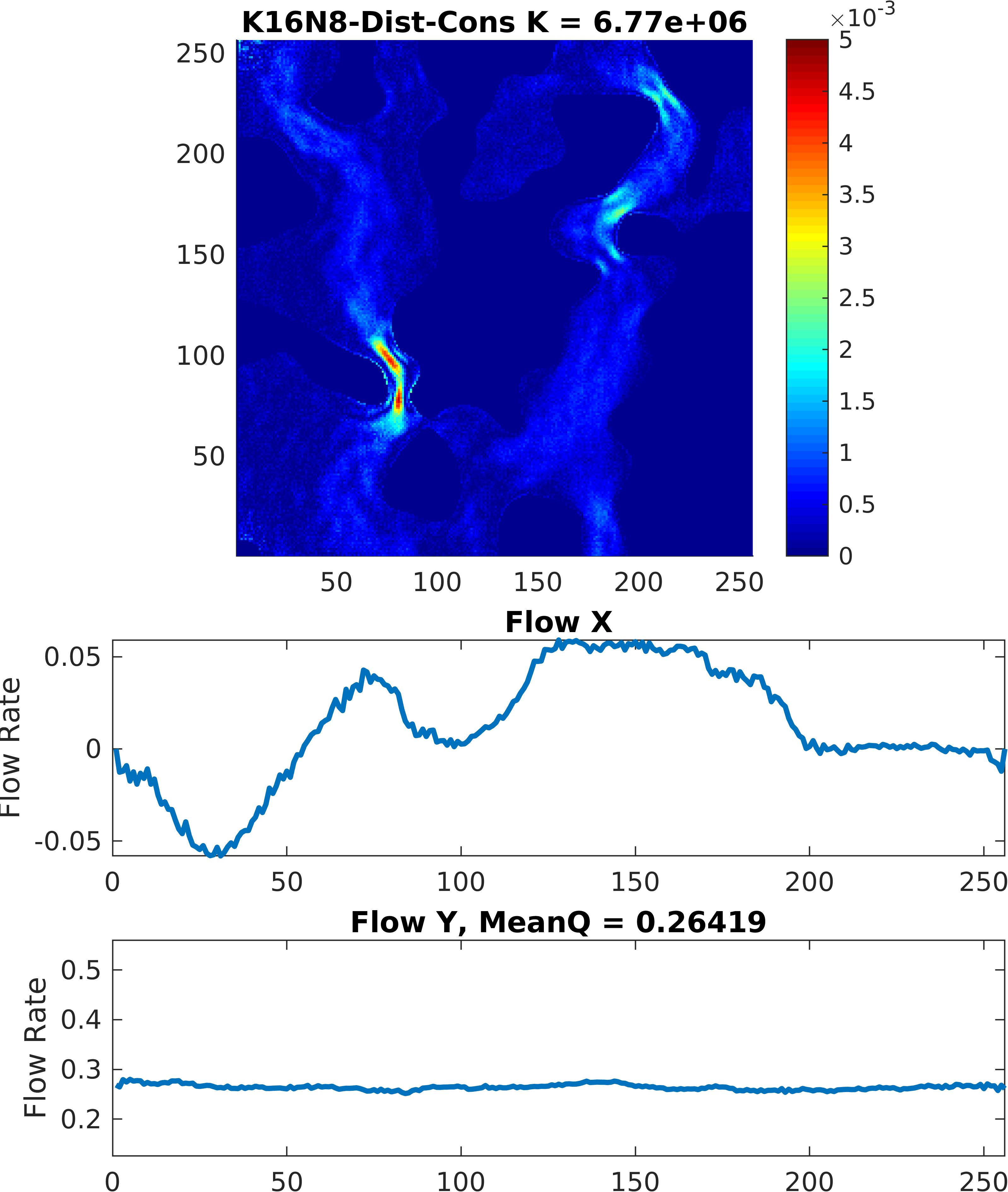}
  \end{minipage}
  \vspace{5mm}
    \begin{minipage}[b]{0.32\textwidth}
    \includegraphics[width=\textwidth]{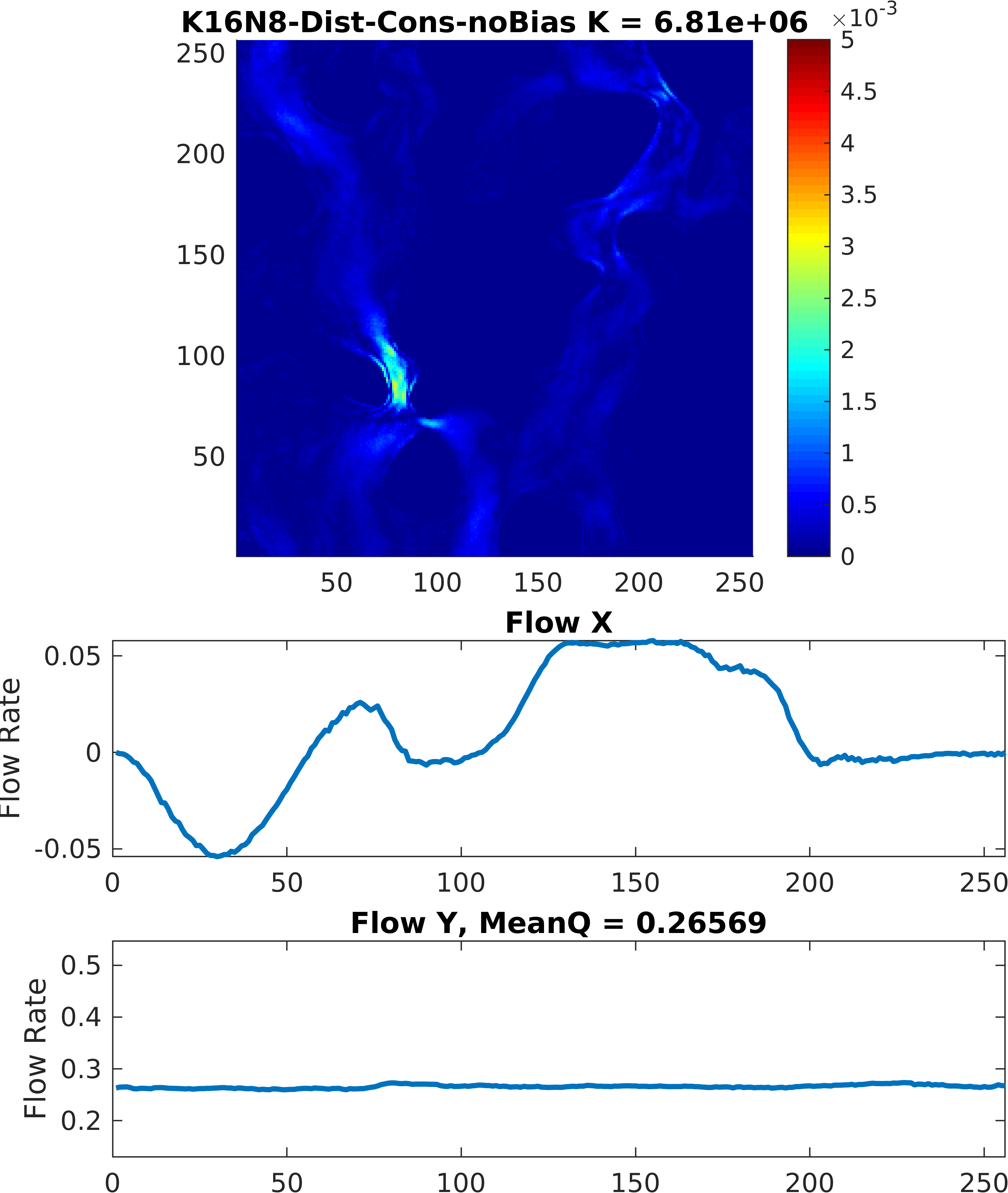}
  \end{minipage}
  \vspace{5mm}
    \caption{From top left to bottom right, the LBM ground truth velocity field, followed by difference maps and cross-sectional comparisons of the predicted velocity fields generated by the various configurations on the median testing image. Difference maps suggest good visual match, also supported by flow rate profiles. The small fluctuations in the flow profiles can be a source of error when using these predicted velocity fields in fine-scale analysis.}
    \label{fig:velComparemedian}
\end{figure}

To illustrate the improvements achieved by these modifications to the network, visual comparisons of the difference map between the predicted velocity fields and the real LBM velocity fields are shown in Figure \ref{fig:velComparemedian}. Plots of the perpendicular mass flux are also shown, and it can be seen that the improvement in accuracy is consistent with trends shown in previous Figures \ref{fig:configCompare} and \ref{fig:configCompareDist}.  

Visually, difference maps show that most errors occur in regions of high velocity, while plots of the cross-sectional mass flux show that, on a slice-by-slice basis, errors manifest as undifferentiable jumps and cusps in the flow profile. While the overall permeability error is low and the velocity fields difference maps suggest a low visual error, the flow rate errors are an indication that the individual voxelwise velocities are misaligned, which is analysed in further detail in a later section. 

\subsection{Effect of Geometric Complexity on Network Accuracy}
\label{sec:complexAccuracy}

\begin{figure}[htp!]
  \centering
    \includegraphics[width=\textwidth]{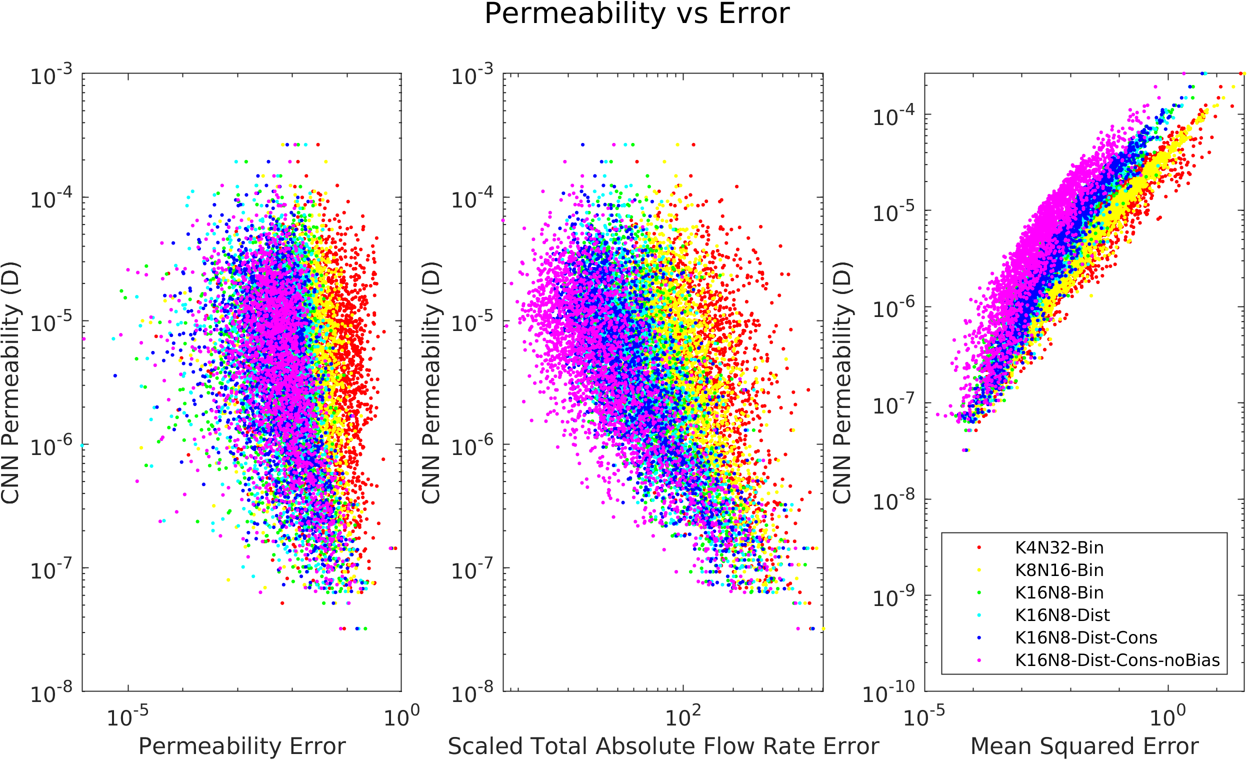}
    \caption{Plots of the permeability of the testing samples vs the various measures of accuracy used in this study for all tested network configurations. A lower permeability (higher geometric complexity) correlates strongly with a higher error in velocity field prediction as measured by the permeability error itself, and the STAFE. MSE errors actually trend towards lower values for lower permeabilities, as the velocity magnitudes in such samples is inherently lower for the same pressure drop}
    \label{fig:permError}
\end{figure}

From plots of the accuracy achieved by the trained networks in previous Figures \ref{fig:configCompare} and \ref{fig:configCompareDist}, a clear distribution of accuracy can be seen. It can be expected that this distribution is due to the variation in the porous structures trained and tested on, so this is quantitatively investigated. Plotting the permeability of each testing sample against the CNN accuracy achieved by the best performing network as shown in Figure \ref{fig:permError} shows very clear trends in the accuracy and the permeability. For the permeabiltiy error (which can be thought of as a mean velocity error) and the STAFE, a lower permeability results in a higher error, suggesting a link between geometric complexity and prediction accuracy. This is not the case for the plot of permeability vs MSE, which tends to higher accuracy for lower values of permeability due to the lower velocity magnitudes predicted in more complex geometries.

\begin{figure}[htp!]
  \centering
     \begin{minipage}[b]{0.5\textwidth}
    \includegraphics[width=\textwidth]{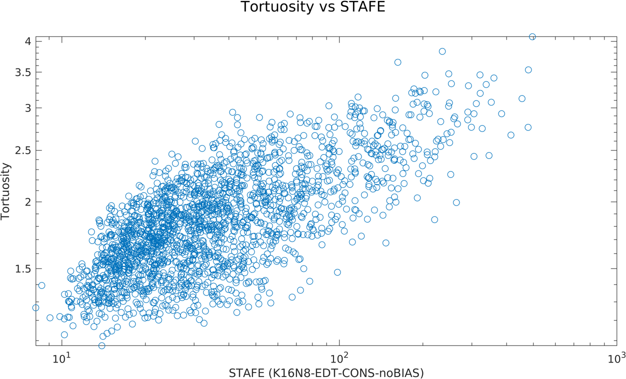}
    \end{minipage}
    \caption{Example 2D images of porous media generated from correlated fields, with varying correlation length, and the Euclidean Distance Transform shown below.}
    \label{fig:tortError}
\end{figure}

\begin{figure}[htp!]
  \centering
  \begin{minipage}[b]{0.49\textwidth}
    \includegraphics[width=\textwidth]{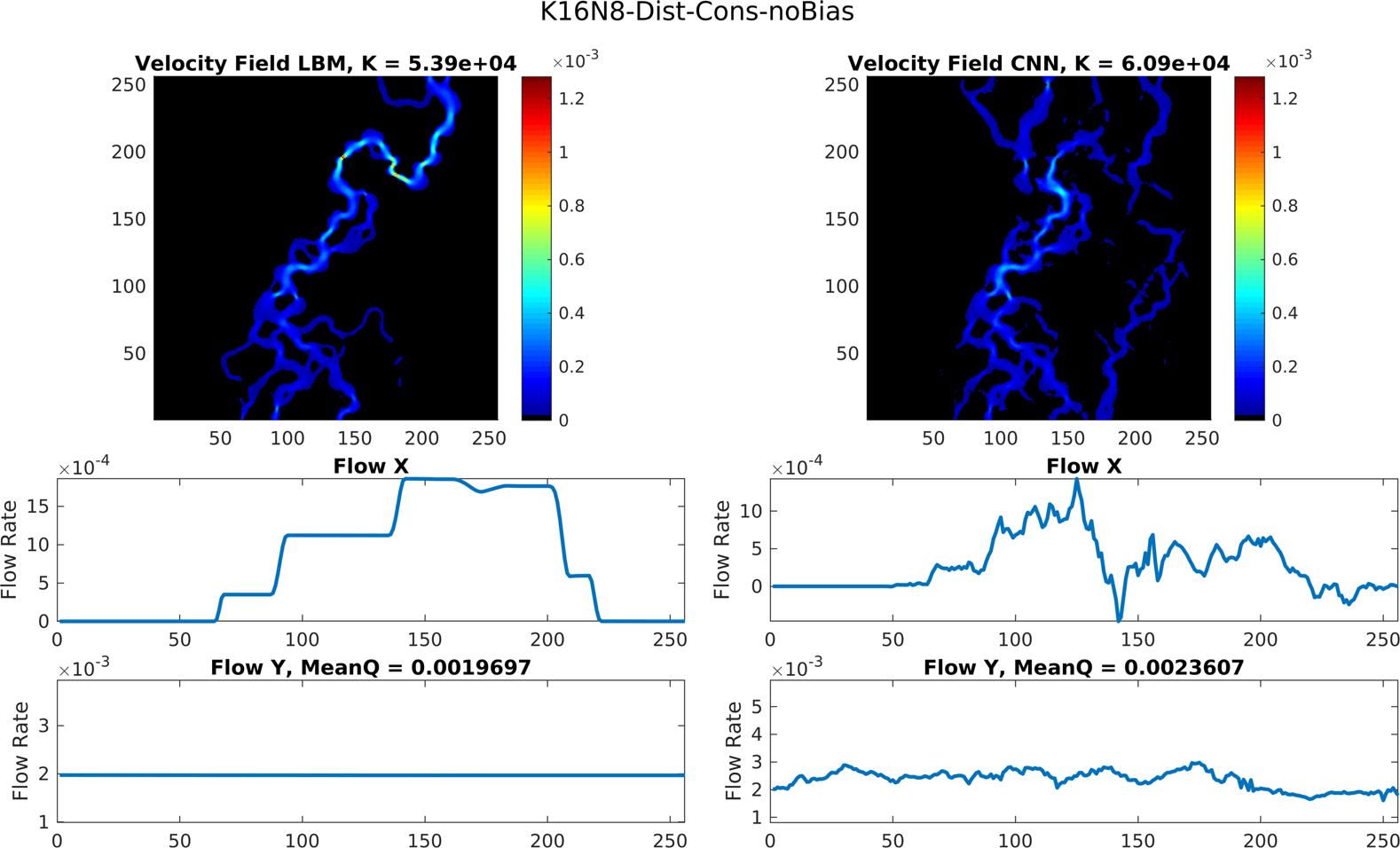}
  \end{minipage}
  \hfill
  \begin{minipage}[b]{0.49\textwidth}
    \includegraphics[width=\textwidth]{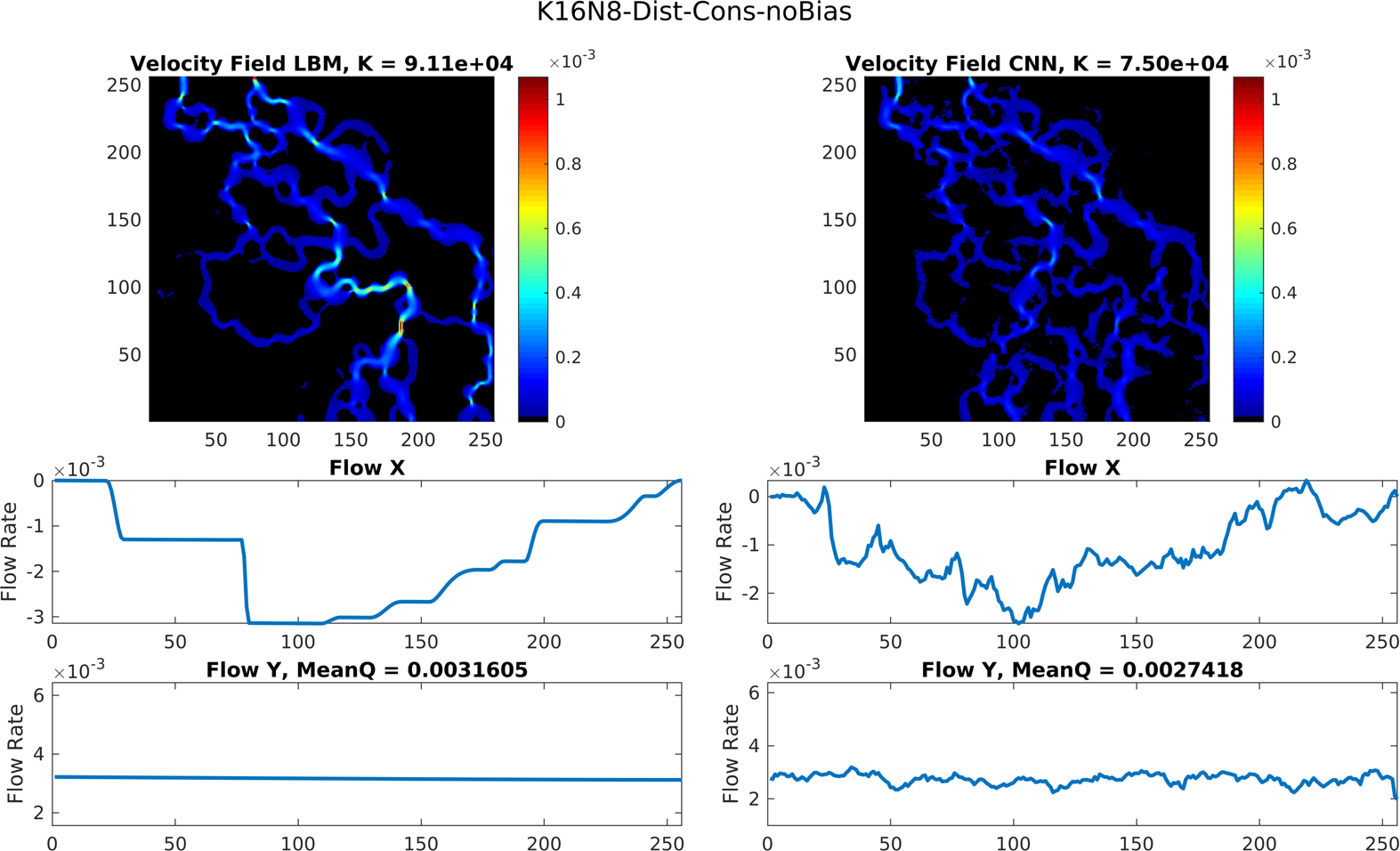}
  \end{minipage}
    \begin{minipage}[b]{0.49\textwidth}
    \includegraphics[width=\textwidth]{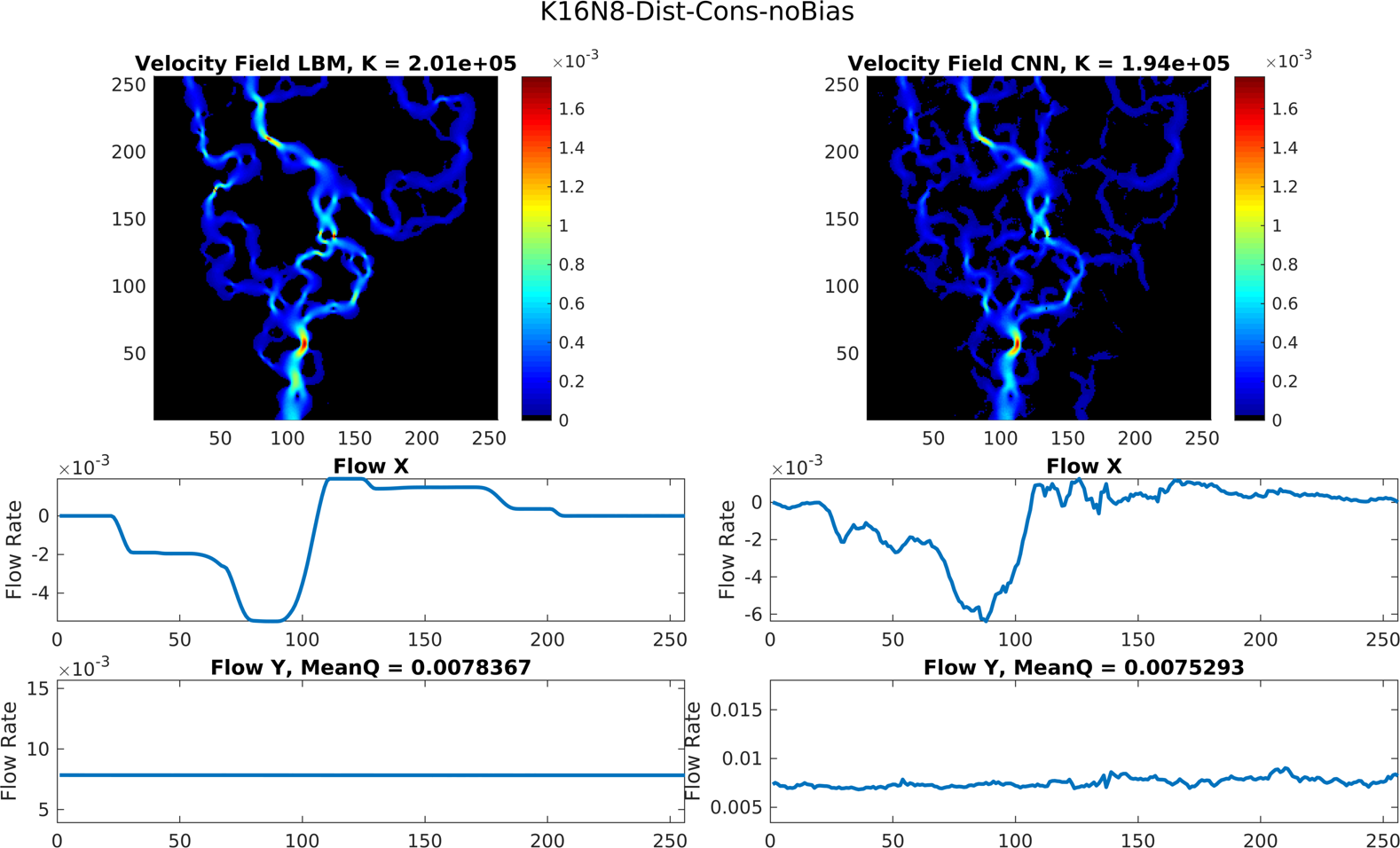}
  \end{minipage}
  \hfill
  \begin{minipage}[b]{0.49\textwidth}
    \includegraphics[width=\textwidth]{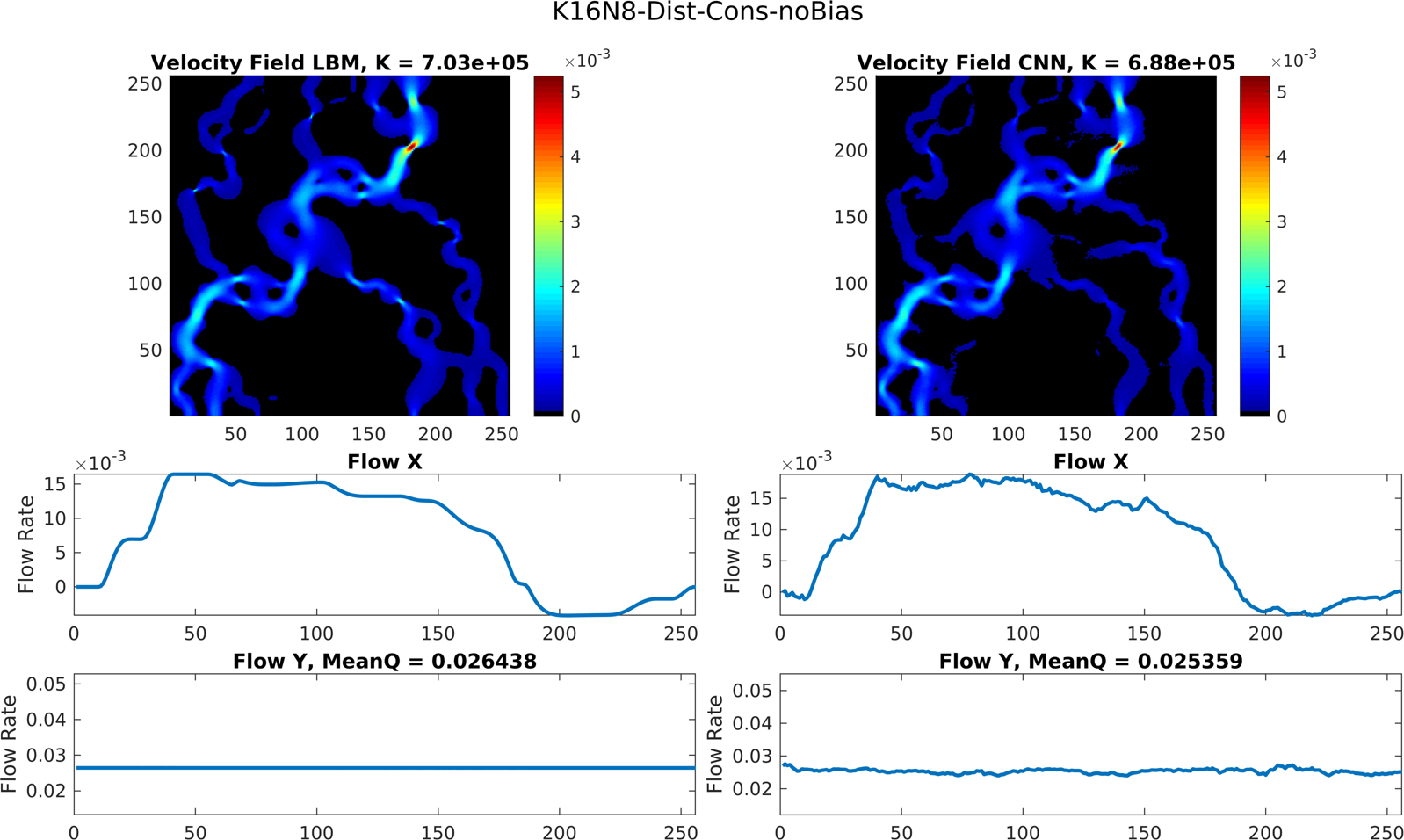}
  \end{minipage}
    \caption{From top left to bottom right, comparison of the predicted velocity fields generated by K16N8 for the worst, 0.5\%, 5\% and 10\% ranked samples based on STAFE metrics within the validation dataset. Overall errors drop dramatically as the geometry is simplified. The increase in flow path width, and the reduction in connected pathways are some factors that can be visually seen to influence this effect}
    \label{fig:velComparek16n8bin}
\end{figure}

To now further visualise this effect, a finer analysis of the velocity fields is required. A plot of the worst, 0.5\%, 5\% and 10\% ranked samples based on the STAFE (a selection of the most complex domains), and the worst, 33\%, 66\% and best ranked samples as generated by the best performing K16N8 network (trained on EDT inputs, with conservation loss, and no biases) on the testing set is shown in Figures \ref{fig:velComparek16n8bin} and \ref{fig:velComparemedian}. These plots reveal that the overall errors drop dramatically once a certain threshold in the domain complexity is reached. The number of possible flow paths and the width of the pore channels are clearly evident factors that influence the performance of the network. These plots show not only that the velocity field prediction is highly dependent on the geometry of the domain, they also show that mass conservation is not enforced adequately, with small to large fluctuations in the slice-by-slice mass flux in the X and Y axes. While the overall error is visually and quantitatively small (see the X and Y difference maps), mass flux profiles are as a whole reasonably predicted, and the velocity distributions are closely aligned, this fluctuation is likely to cause problems in directly using these velocity fields for other applications. 

\begin{figure}[htp!]
  \centering
  \begin{minipage}[b]{0.49\textwidth}
    \includegraphics[width=\textwidth]{figures/velCNNs494-6.png}
  \end{minipage}
  \hfill
  \begin{minipage}[b]{0.49\textwidth}
    \includegraphics[width=\textwidth]{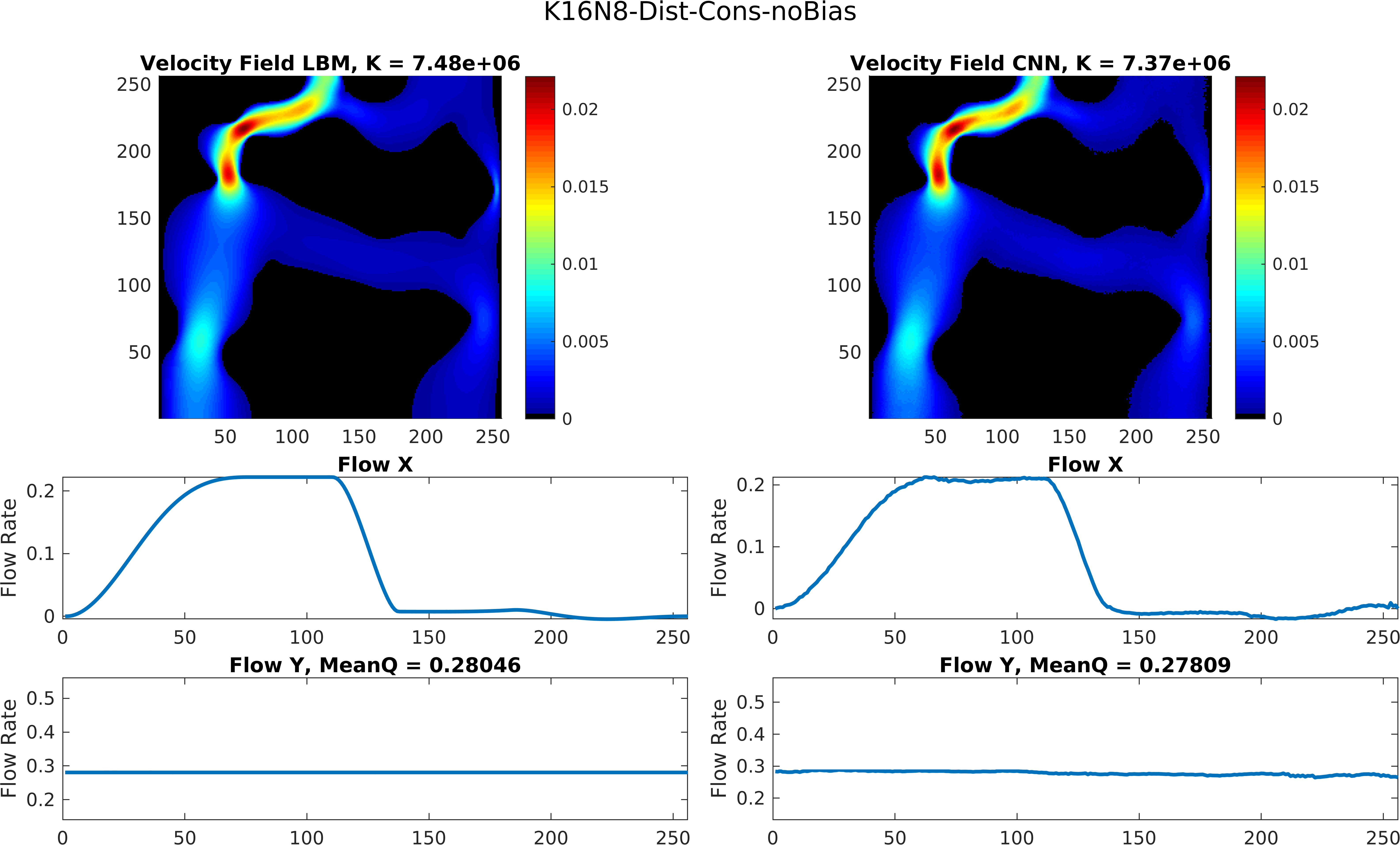}
  \end{minipage}
    \begin{minipage}[b]{0.49\textwidth}
    \includegraphics[width=\textwidth]{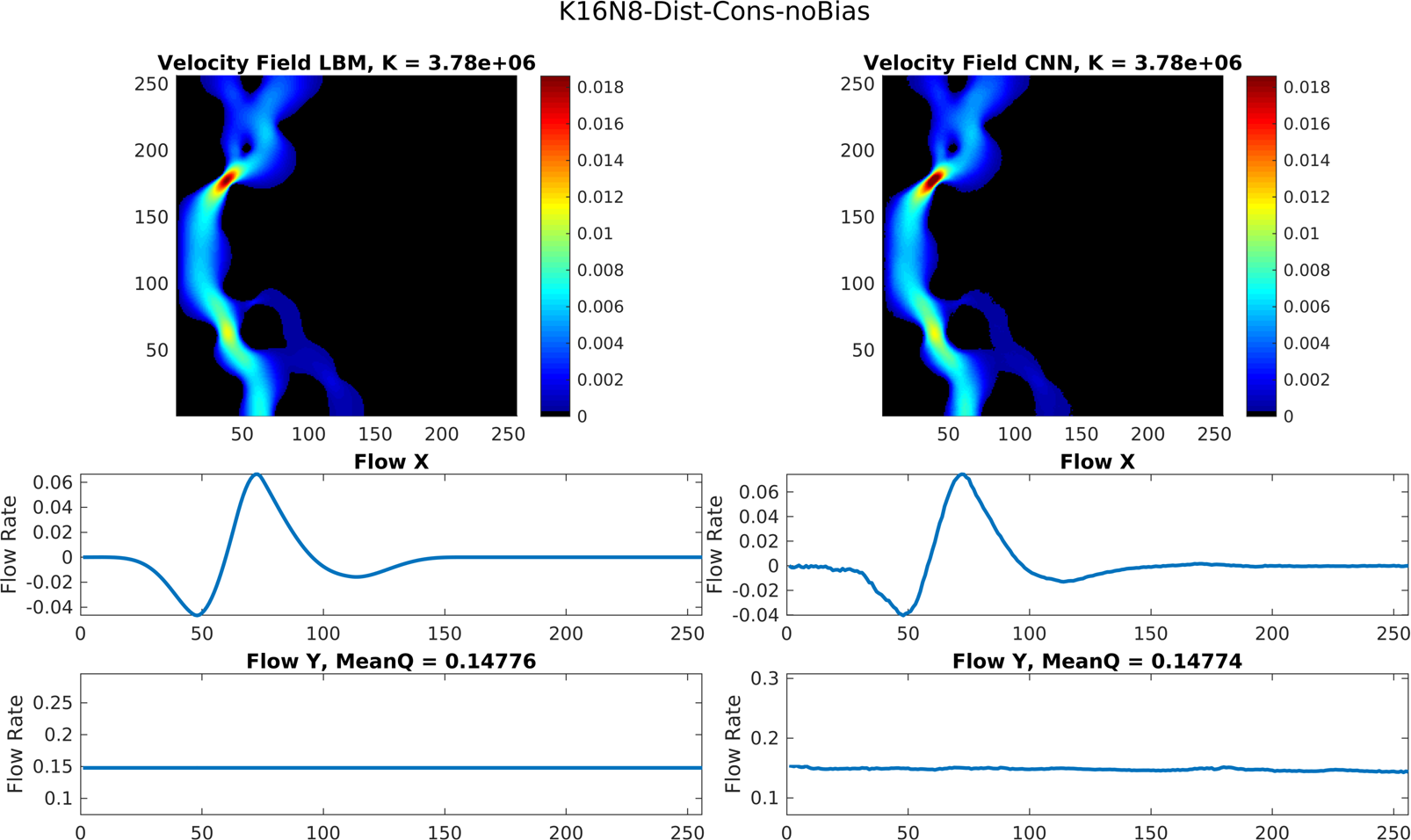}
  \end{minipage}
  \hfill
  \begin{minipage}[b]{0.49\textwidth}
    \includegraphics[width=\textwidth]{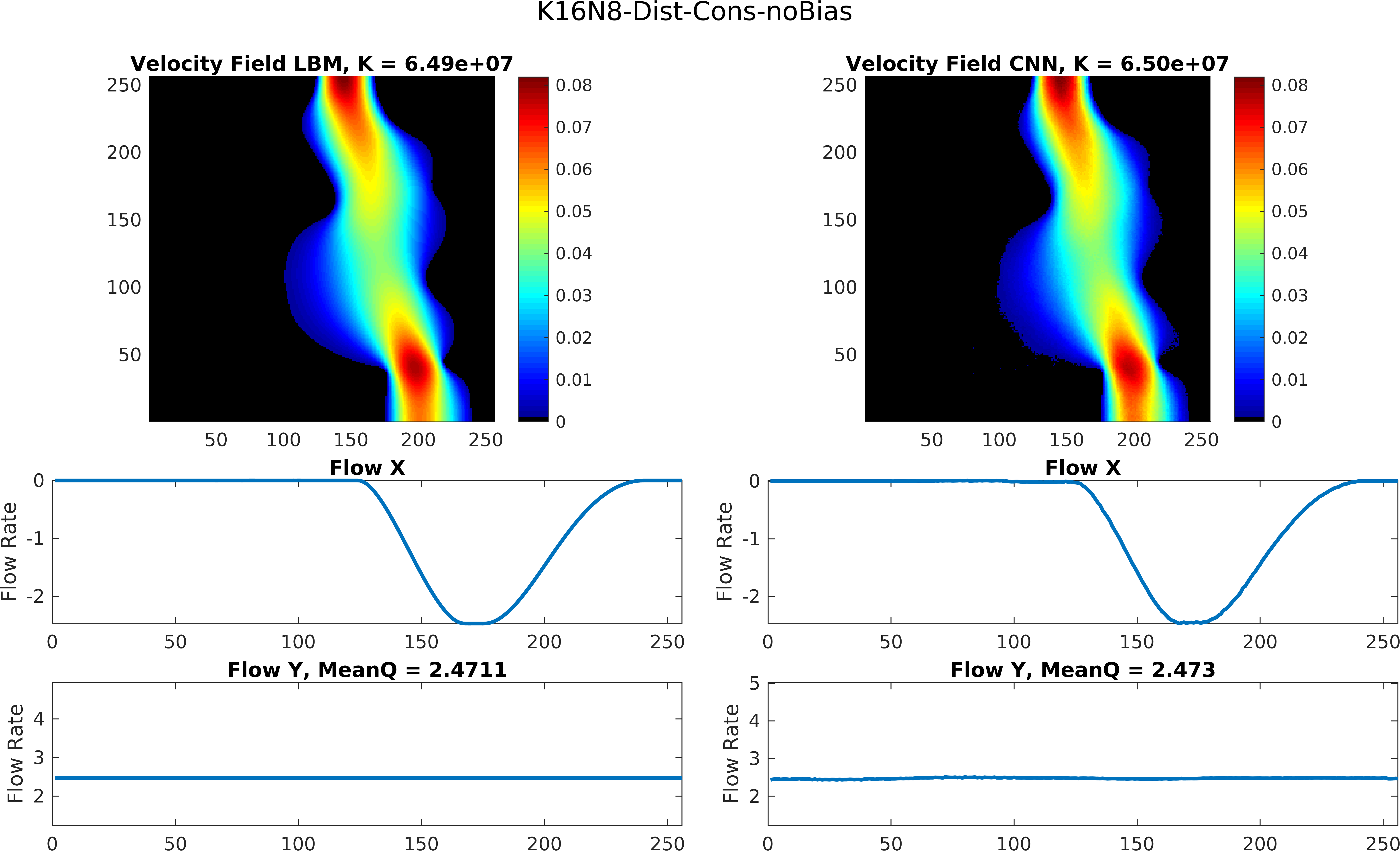}
  \end{minipage}
    \caption{From top left to bottom right, comparison of the predicted velocity fields generated by K16N8 for the worst, 33\%, 66\% and best samples based on STAFE metrics within the validation dataset. There is a clear trend, were geometric complexity results in higher errors. Lower error samples have larger, smoother, and wider flow channels. It can be expected that, CNN performance is acceptable up to a certain degree of complexity, and should be applied in a local-to-global manner}
    \label{fig:velComparek16n8binworst}
\end{figure}

As can be seen, there is a clear relationship between the accuracy achieved in predicting the velocity field and the geometric complexity of the domain. The best matches occur in relatively simple tubular structures without any branching pathways, while the worst matches occurs in domains with multiple thin, low resolution pathways. When predicting the velocity field of a domain, these limitations should be considered. While a patch based approach may be applied whereby subsections of an overall domain are fed into the network, the local prediction of velocity fields is inconsistent with the highly non-local true solution. A wide flow path outside the bounds of a local patch will significantly affect local velocity fields. This issue can possibly be addressed by encoding some information regarding the magnitude of the local velocity field relative to the unseen global domain. In the interests of preserving accuracy however, a fully global reconstruction methodology is recommended. U-net and its variants are fully convolutional (meaning that the trained network can be applied to domains of any size), so domain size is variable and useful if the problem is non-local such is the case with flow as it is dependent on boundary conditions and far-away geometries.

\subsection{Permeability Estimation}
\label{sec:permest}
\begin{figure}[htp!]
    \includegraphics[width=0.49\textwidth]{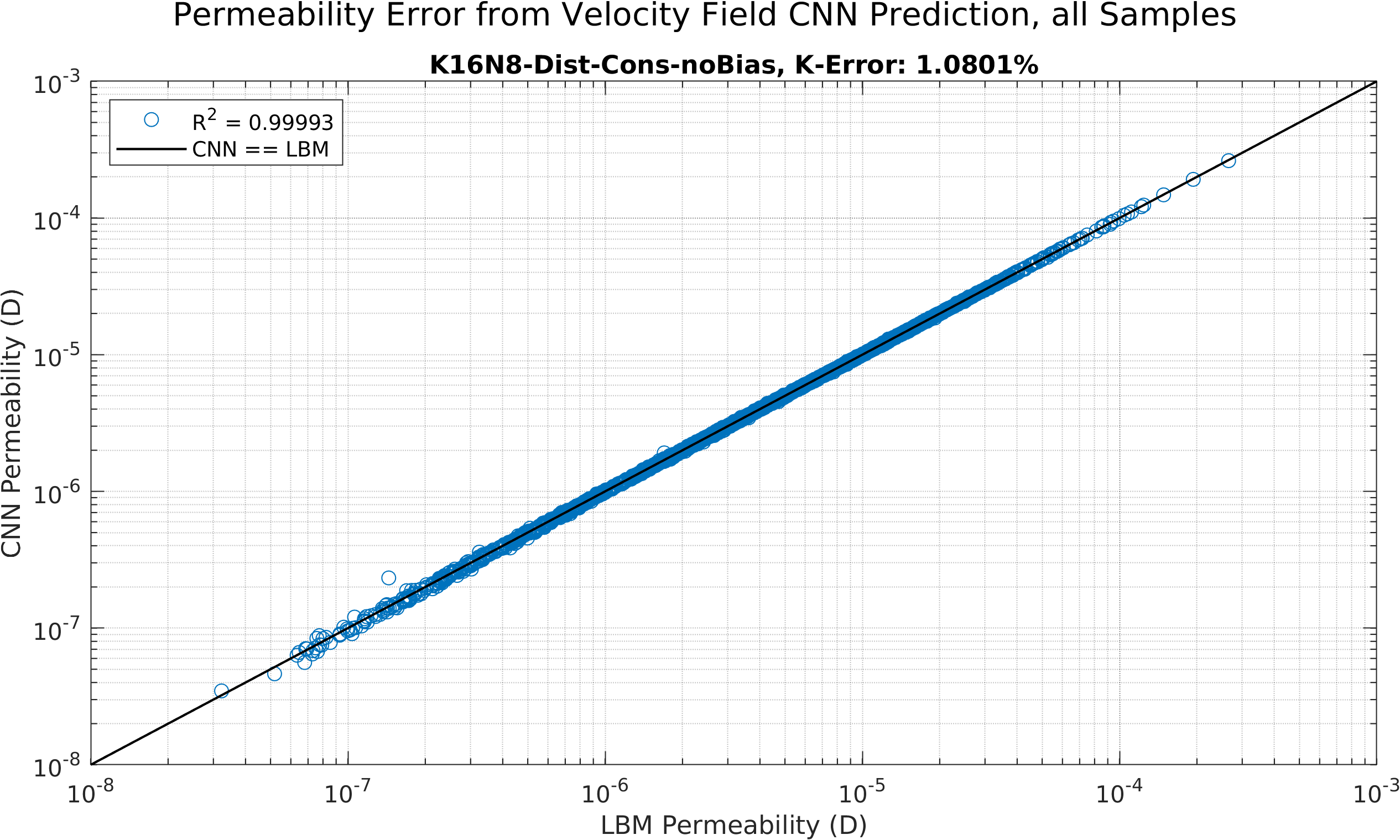}
    \includegraphics[width=0.49\textwidth]{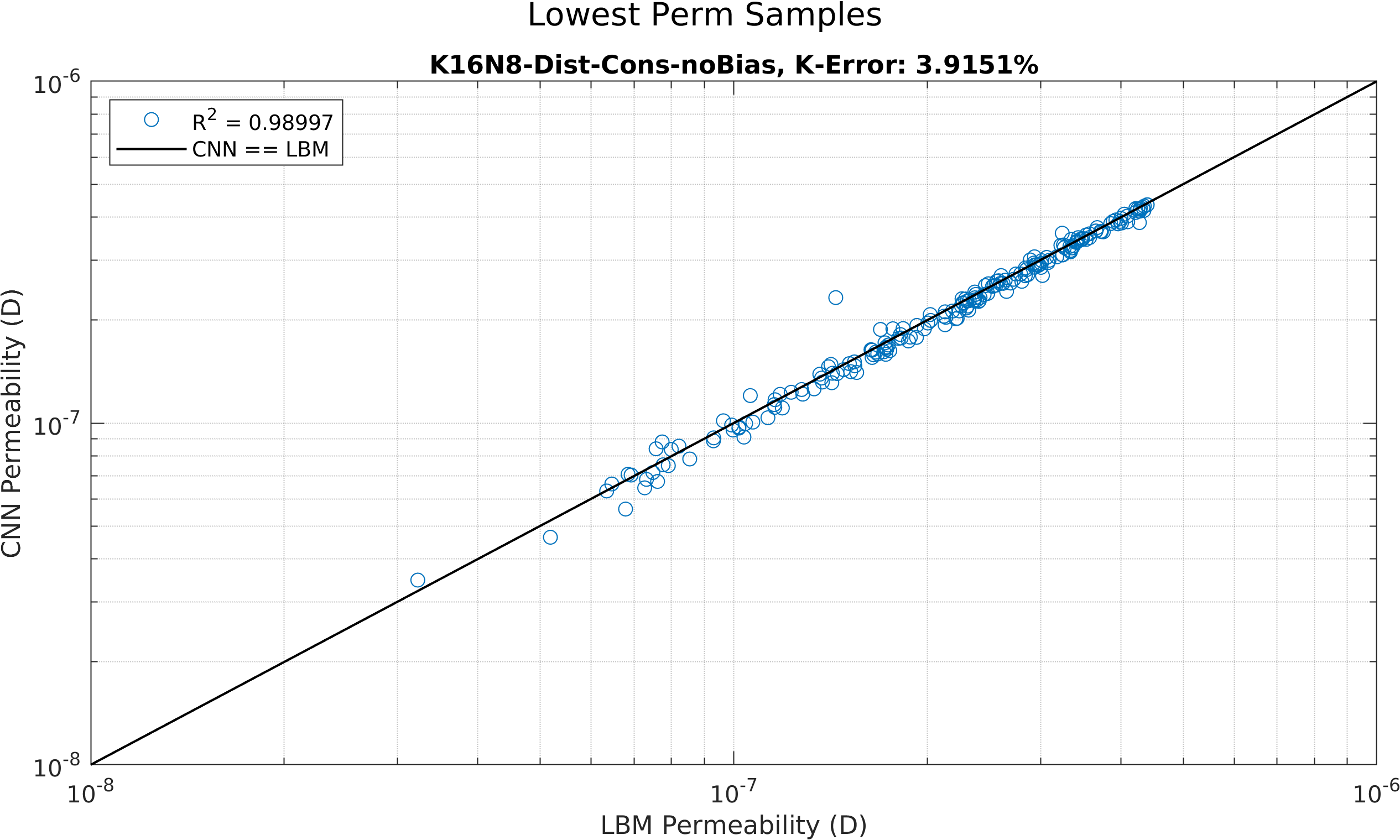}
    \caption{Permeability prediction accuracy of velocity field predictions. Overall errors are low, even though local velocity field error is high, seen in previous sections}
    \label{fig:permpermError}
\end{figure}

In contrast to predicting permeability by regression, whereby a single number is directly obtained by a model such as a Neural Network, the prediction of permeability by first estimating the velocity fields presents finer, interpretable detail based on the geometric complexities of the domain. As has already been shown in Figures \ref{fig:configCompare}, \ref{fig:configCompareDist}, and \ref{fig:permError}, overall errors are low, with 99\% falling below 10\% and 80\% falling below 1\% error. This is depicted in a more traditional plot of real vs predicted permeability, shown in Figure \ref{fig:permpermError} of all 2,000 unique 2D testing images spanning 4 orders of magnitude, and of the permeability error of the 100 samples with the lowest permeability (and highest error). The accuracy and flexbility shown by the predictive model is on par with regression based CNN models. An important point to note is the relative insensitivity of permeability prediction compared to the fine-scale flow and velocity predictions. In plots of the 100 lowest permeability testing samples, the average permeability error for the best performing network configuration remains low, under 4\% for the most geometrically complex testing samples. This does not translate well to the errors associated with velocity fields and flow profiles, as shown in Section \ref{sec:complexAccuracy}. Both regression and velocity prediction based methods tend to give good estimates of the permeability of porous media. These 2D results are expected to be more accurate than those achievable in 3D, as the topology of the media is important for the prediction and this is more complex in 3D. This is explored in Section \ref{sec:3Dnetwork}

Another important point to note is that, since the velocity sensitivity is much higher than permeability sensitivity, for the purposes of measuring the accuracy achieved by velocity field prediction, the permeability is a poor candidate. It is common for permeability errors in the order of 10\% to be considered a good result using semi-analytical models, pore network models, or otherwise \cite{pfvs}, but in this case, we see that even a 1\% error (or significantly less) in the permeability as shown in Figure \ref{fig:permpermError} results in a significant deviation in the locally predicted velocity fields as shown in Figures \ref{fig:velComparek16n8bin} and \ref{fig:velComparek16n8binworst}. To better understand the fine-scale errors, plots of the cross-sectional flow profiles, and the STAFE accuracy measure are better representations of velocity prediction. 

\begin{figure}[htp!]
  \centering
    \includegraphics[width=0.8\textwidth]{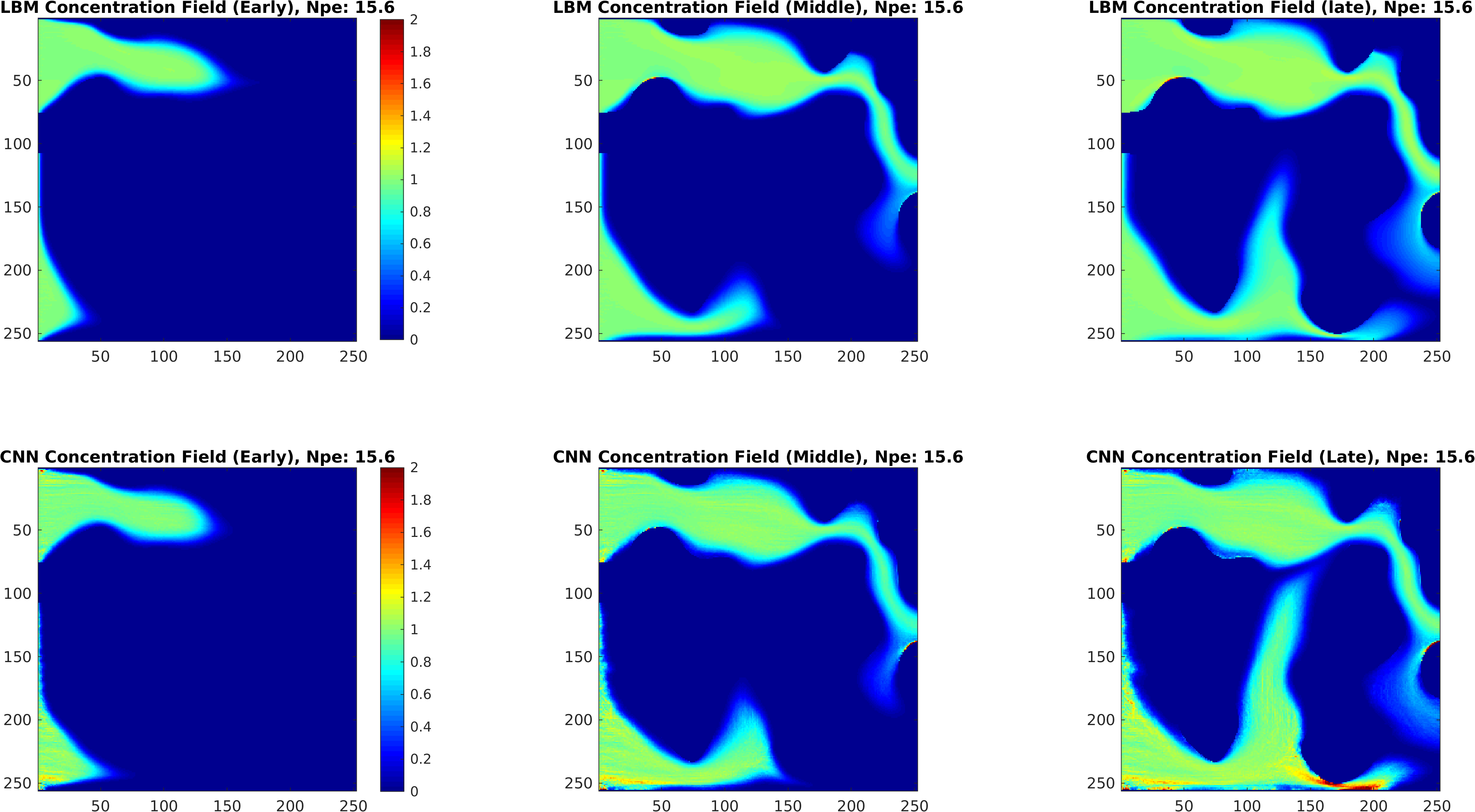}
  \vspace{5mm}
      \includegraphics[width=0.8\textwidth]{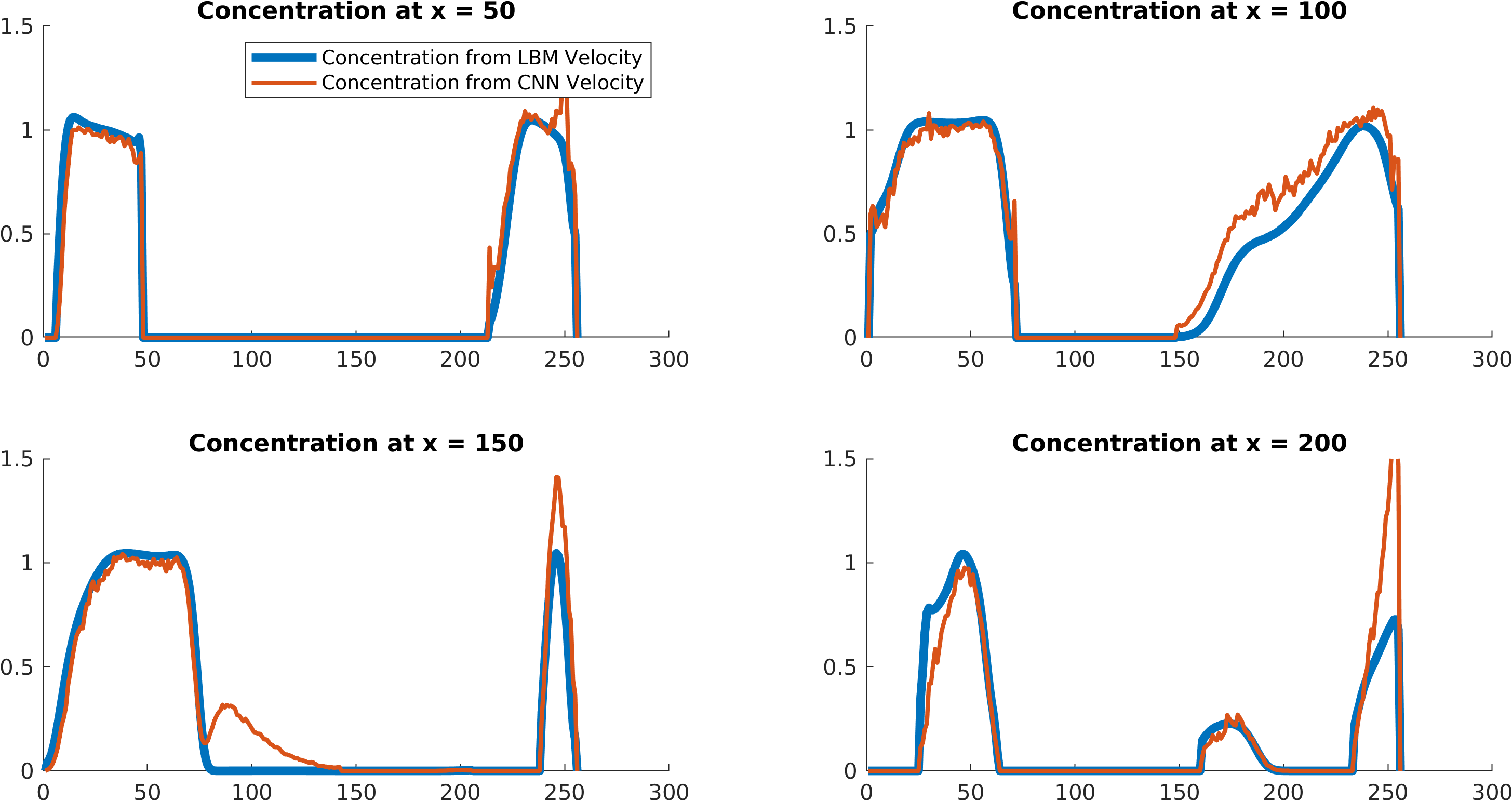}
    \caption{Top: Comparison of concentration fields generated by convection diffusion with underlying velocity fields sourced from both LBM and CNN predictions. The fine-scale misalignment and errors in the CNN velocity fields at the walls and at throats can be seen to cause errors in the form of mass accumulation. Bottom: Cross section comparison of concentration profiles at Late time. The CNN results show erroneous accumulation and irregular peaks due to misalignment of velocity field vectors from CNN prediction.}
    \label{fig:velCompareConc}
\end{figure}

\subsection{Fine Scale Sensitivity of Solute Transport to Predicted Fields}
\label{sec:2Dconvdiff}
Another such test that can highlight the problematic errors that occur in velocity field prediction is the use of such velocity fields for analysis of solute transport in porous media. In this case, 2D convection-diffusion within the pore space is modelled using a Finite Volume solution of the convection diffusion equation. An inlet concentration of 1 is set, and an open outlet boundary condition is set. To highlight the influence of velocity fields, the Peclet Number is set to a reasonably high value, near the limits of computational diffusion. An explicit time-stepping scheme is used, again to minimise numerical diffusion. In order to also ensure stability, upstreaming is applied instead of TVD methods. 

The issue with errors at the fine-scale is shown easily with a single, simple example, as shown in Figure \ref{fig:velCompareConc}, of a relatively well predicted field (the testing sample with 33-percentile accuracy, shown in Figure \ref{fig:velComparek16n8binworst}). While the real velocity fields result in expected transport of solute, without major zones of erroneous accumulation, the misaligned and off-magnitude velocity fields predicted near the wall, and in tight throats clearly shows a high error when applying these fields to sensitive tasks, such as solute transport. This error is shown in further detail in the cross sectional concentration profiles at x = 50, 100, 150, and 200 for the late time concentration fields. While the concentration profiles are reasonably well matching in their shape, the increased noise and the mismatching peaks are indicative of erroneous solute accumulation due to misaligned velocity vectors. This type of error can be problematic in cases where the accumulation of solute is a focal point in analysis, such as contaminant transport modelling.

\begin{figure}[htp!]
  \centering
    \includegraphics[width=\textwidth]{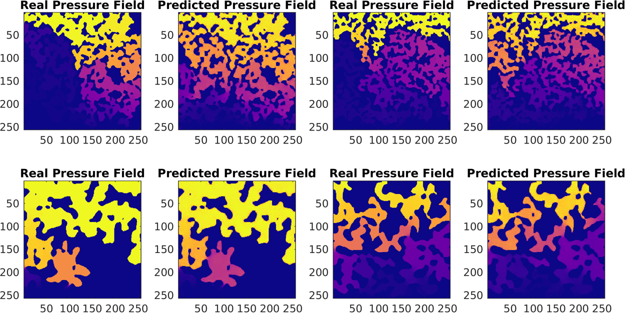}
     \begin{minipage}[b]{0.6\textwidth}
    \includegraphics[width=\textwidth]{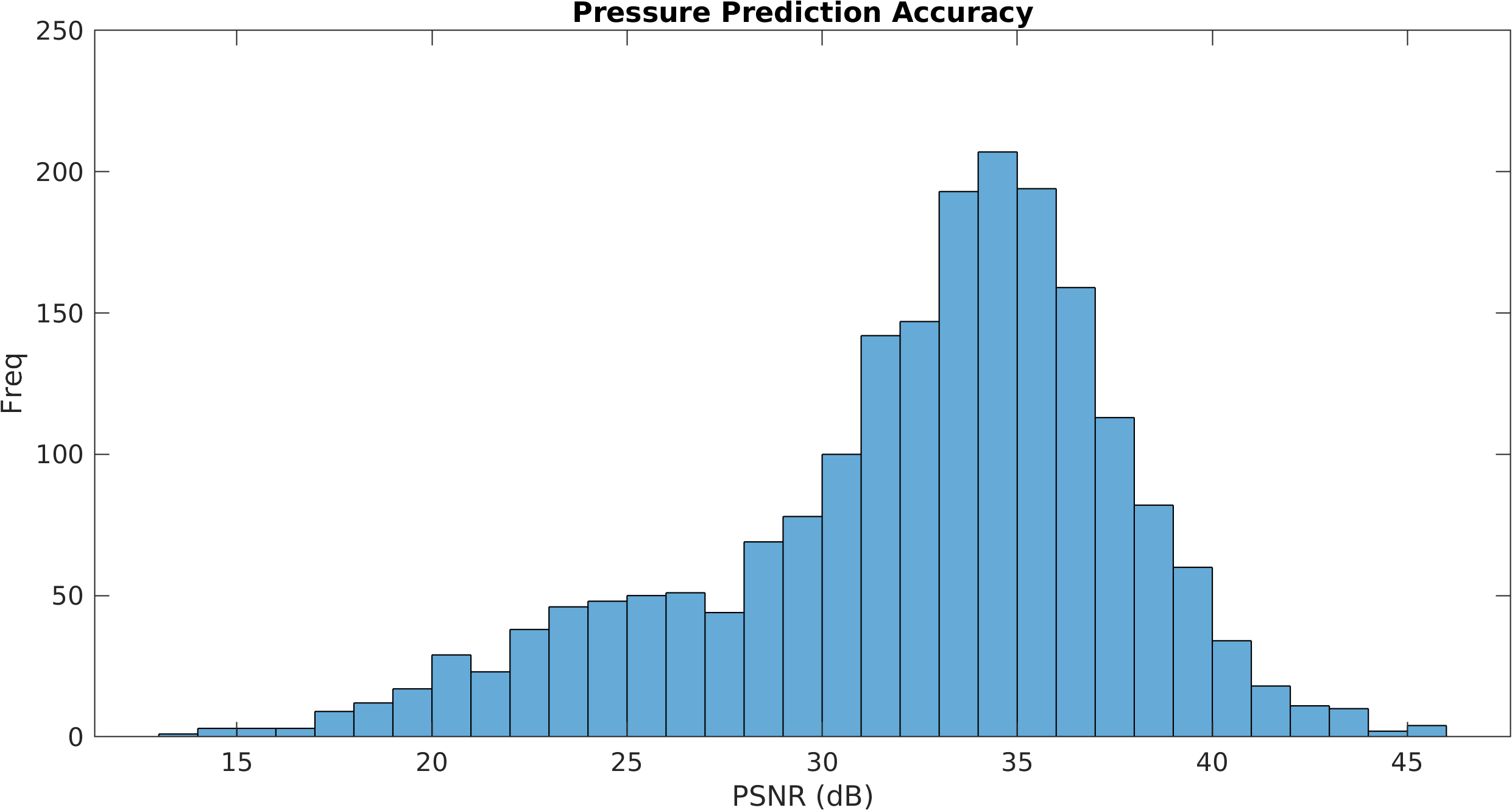}
    \end{minipage}
    \caption{Top: Examples comparing the pressure field obtained from LBM, and the predicted pressure field generated by the network. The major pressure gradients are captured accurately, though there is pressure mismatch in some porous chambers. Bottom: Histogram of the accuracy achieved by pressure field prediction over the testing dataset. With the exception of the most complex and tortuous geometries, accuracy is high, with the majority of cases achieving a PSNR of over 20 dB, corresponding to an L2 error of 1\%}
    \label{fig:pressCNN}
\end{figure}

\subsection{Accelerating Flow Simulation to Steady State}
\label{sec:acc2D}
A potential solution to fix these prediction errors shown in previous sections is to use these velocity field predictions as initial estimates in direct flow simulation to realign the velocity fields and accelerate the computation to steady state. To accomplish this, both the velocity field and pressure fields are required for a given domain. Thus the network is trained to predict both the velocity field as well as the pressure field. Pressure fields obtained from LBM are scaled to [0 1], and samples are shown in Figure \ref{fig:pressCNN} and the resulting accuracy is measured by the Peak Signal-to-Noise Ratio (PSNR) in decibels obtained on predicting the pressure fields of the 2,000 testing images.

The visualisation of the predicted pressure fields shows that pressure prediction is reasonably accurate, though much like the case in velocity field prediction, chambers that are not prone to flow are highlighted by the network, as identification of the principle flow channels remains difficult with this implementation. PSNR plots confirm that the accuracy is relatively high. For ease of reference, a PSNR of 10 corresponds to a MSE of 10\%, a PSNR of 20 is 1\% and so on. 

\begin{figure}[htp!]
  \centering
  \begin{minipage}[b]{0.49\textwidth}
    \includegraphics[width=\textwidth]{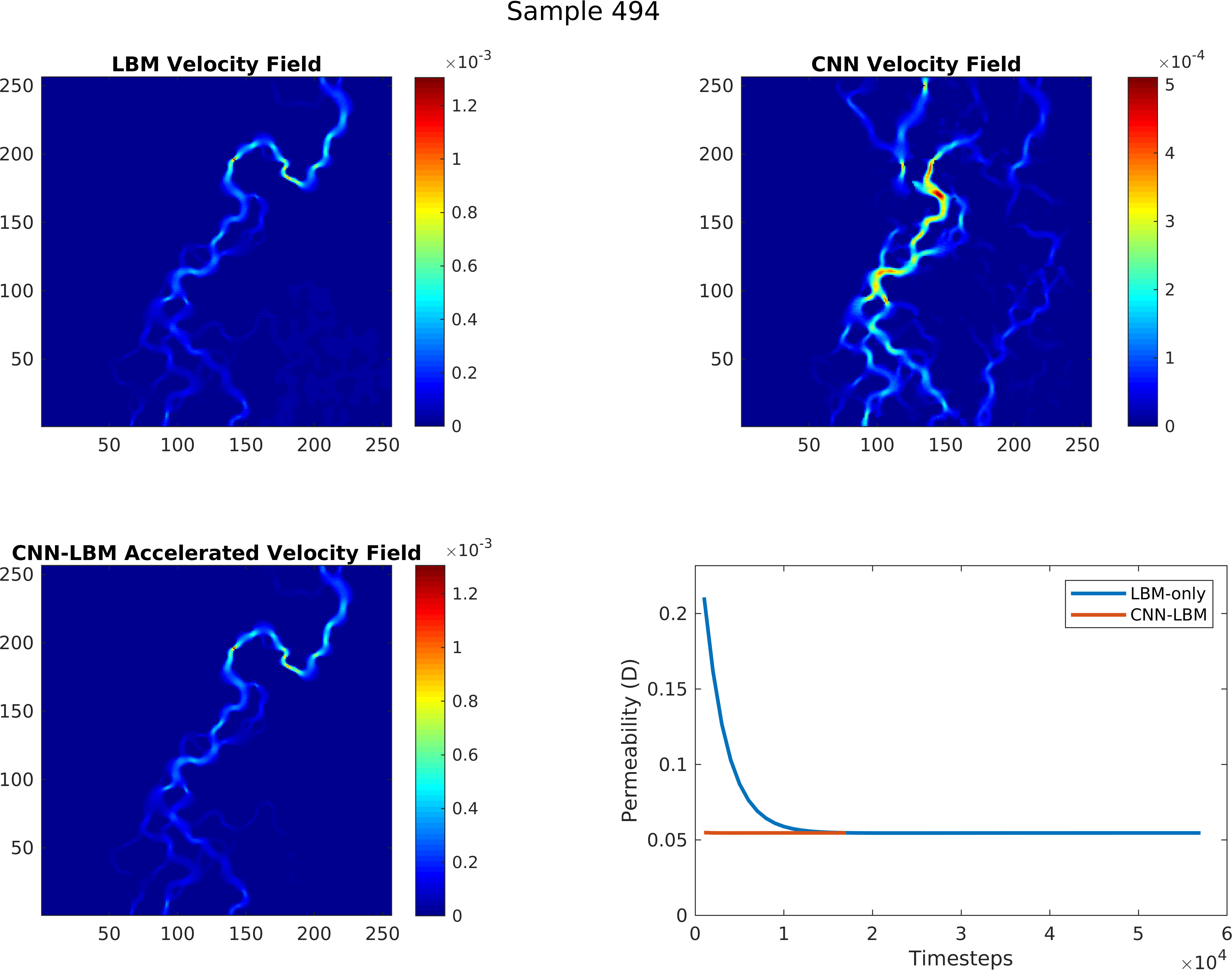}
  \end{minipage}
  \hfill
  \begin{minipage}[b]{0.49\textwidth}
    \includegraphics[width=\textwidth]{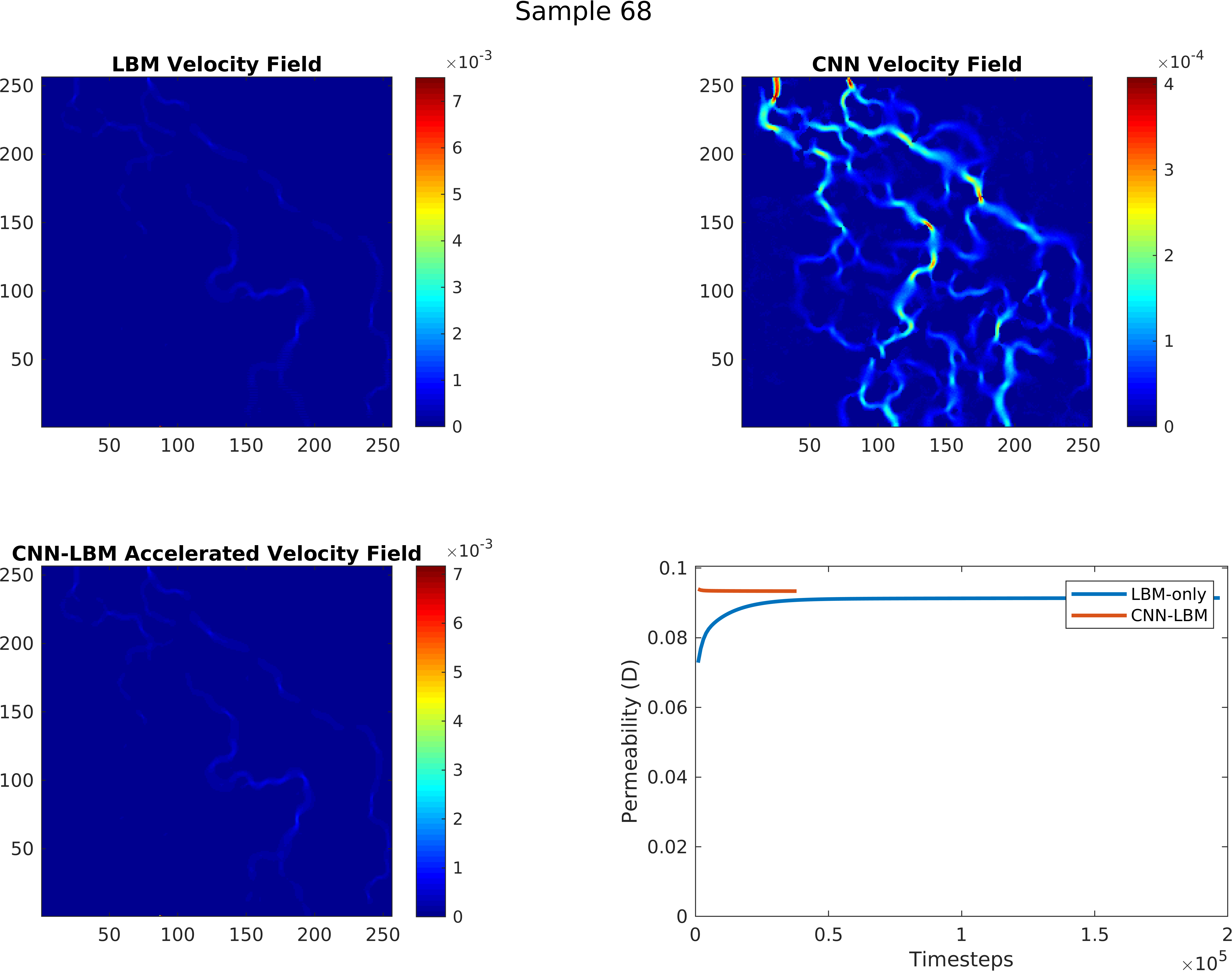}
  \end{minipage}
    \begin{minipage}[b]{0.49\textwidth}
    \includegraphics[width=\textwidth]{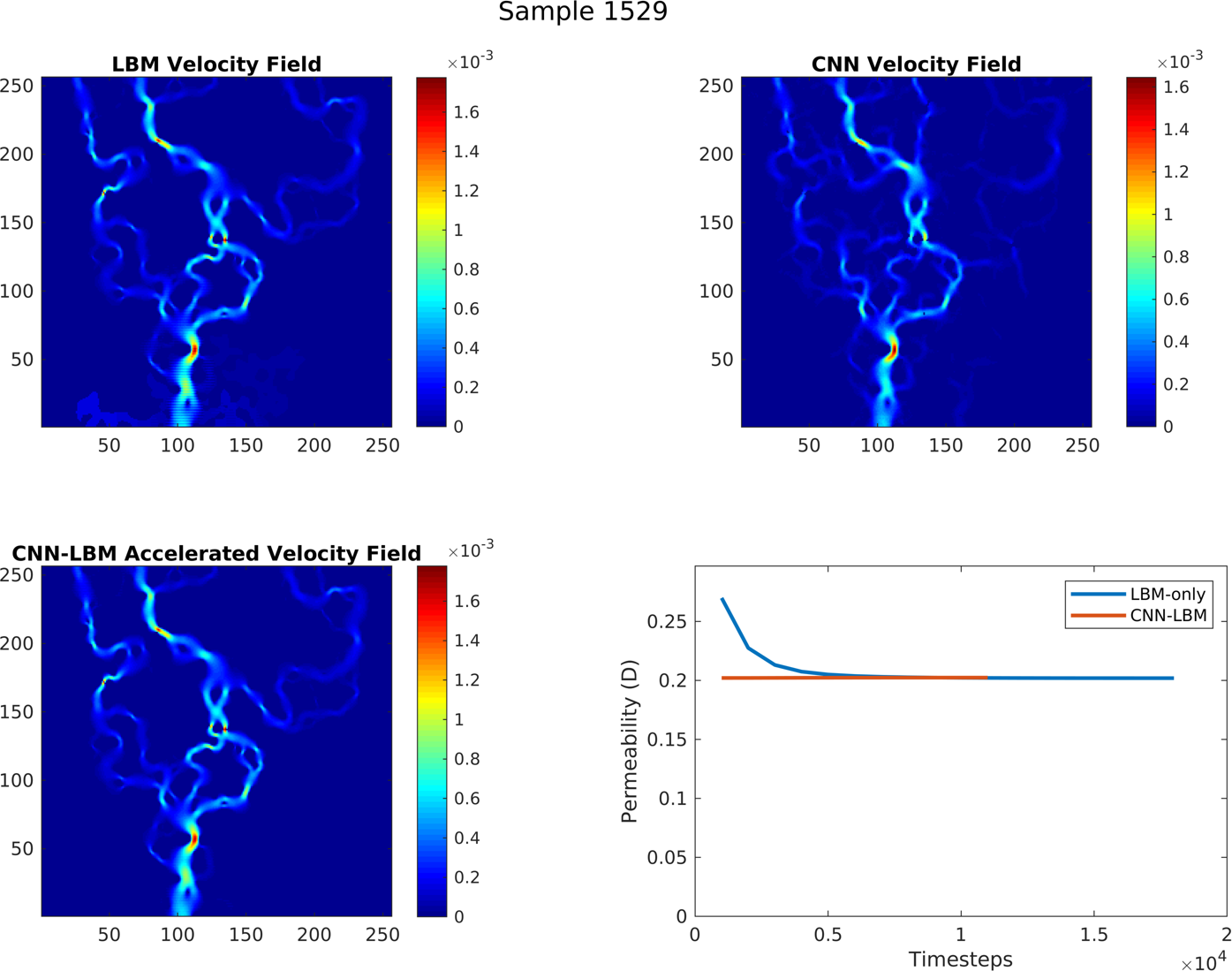}
  \end{minipage}
  \hfill
  \begin{minipage}[b]{0.49\textwidth}
    \includegraphics[width=\textwidth]{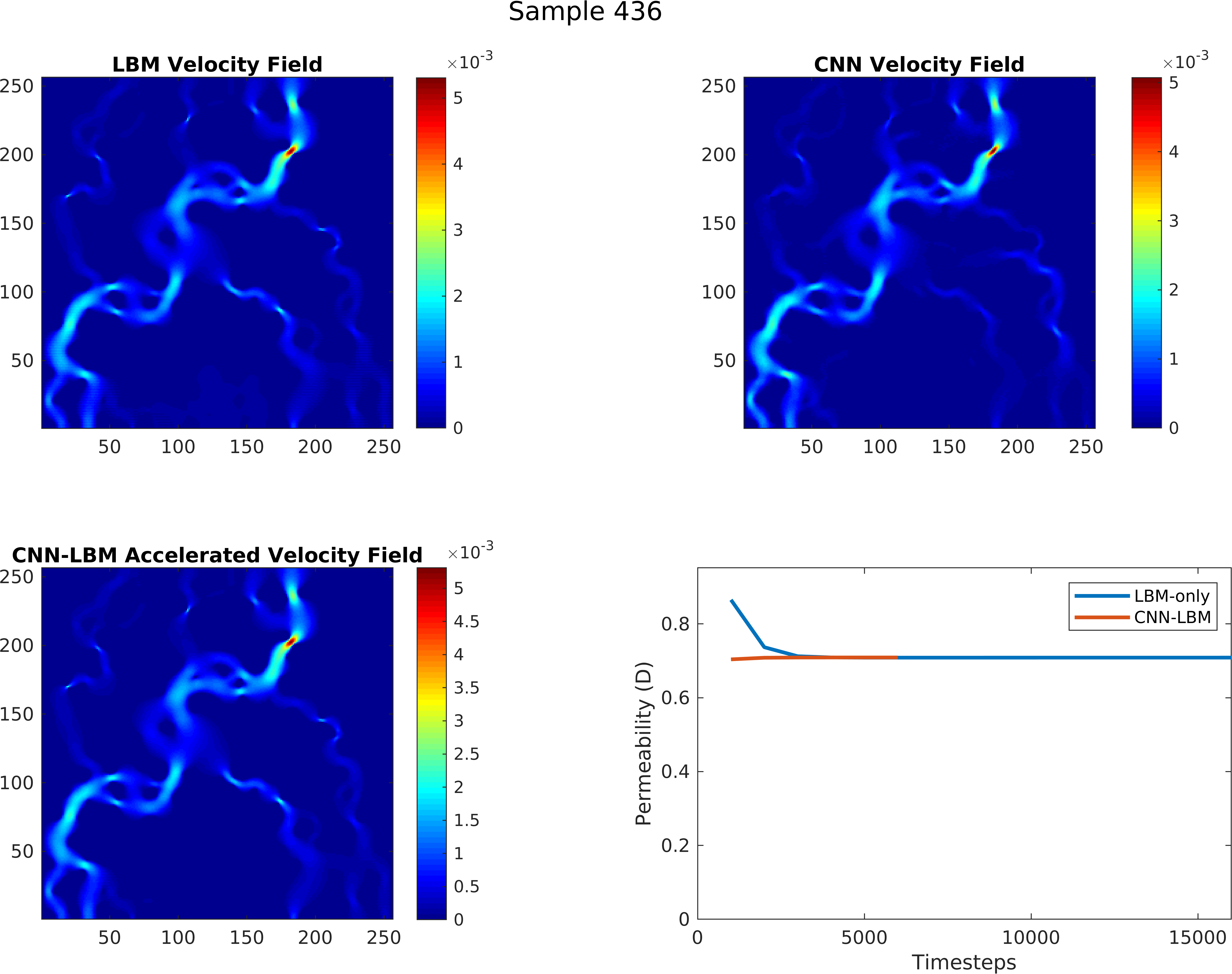}
  \end{minipage}
    \caption{From top left to bottom right, comparison of the LBM-CNN Accelerated velocity fields generated for the worst, 0.5\%, 5\% and 10\% ranked samples within the validation dataset. Even for the worst case, where predicted velocity fields are poorly predicted, the LBM simulation converges rapidly to the true steady state solution. In all cases shown here, representing some of the most error-prone geometries, the benefits of acceleration to steady state is apparent.}
    \label{fig:velAccelerate}
\end{figure}

Using both the velocity predictions and the pressure field predictions as input into LBM (or any direct Navier-Stokes solver), the speed up in convergence to steady state conditions is tracked and compared to the case with simulation initialised with zero velocity and uniform pressure. The convergence criteria is defined in this case by the change in calculated system permeability every 1000 LBM timesteps. In these simulations, this criteria is set to 1e-5. Plots of the worst, 0.5\%, 5\% and 10\% ranked samples are shown in Figure \ref{fig:velAccelerate}, and indicate that the initialisation with even poorly predicted velocity fields results in rapid convergence to steady state conditions. A speed up of 2-5 is observed when terminating simulations at the aforementioned 1e-5 criteria, and resulting velocity fields are identical to those obtained form LBM-only simulations that take longer to stabilise.

\subsection{3D Network Performance}
\label{sec:3Dnetwork}
\begin{figure}[htp!]
  \centering
    \includegraphics[width=\textwidth]{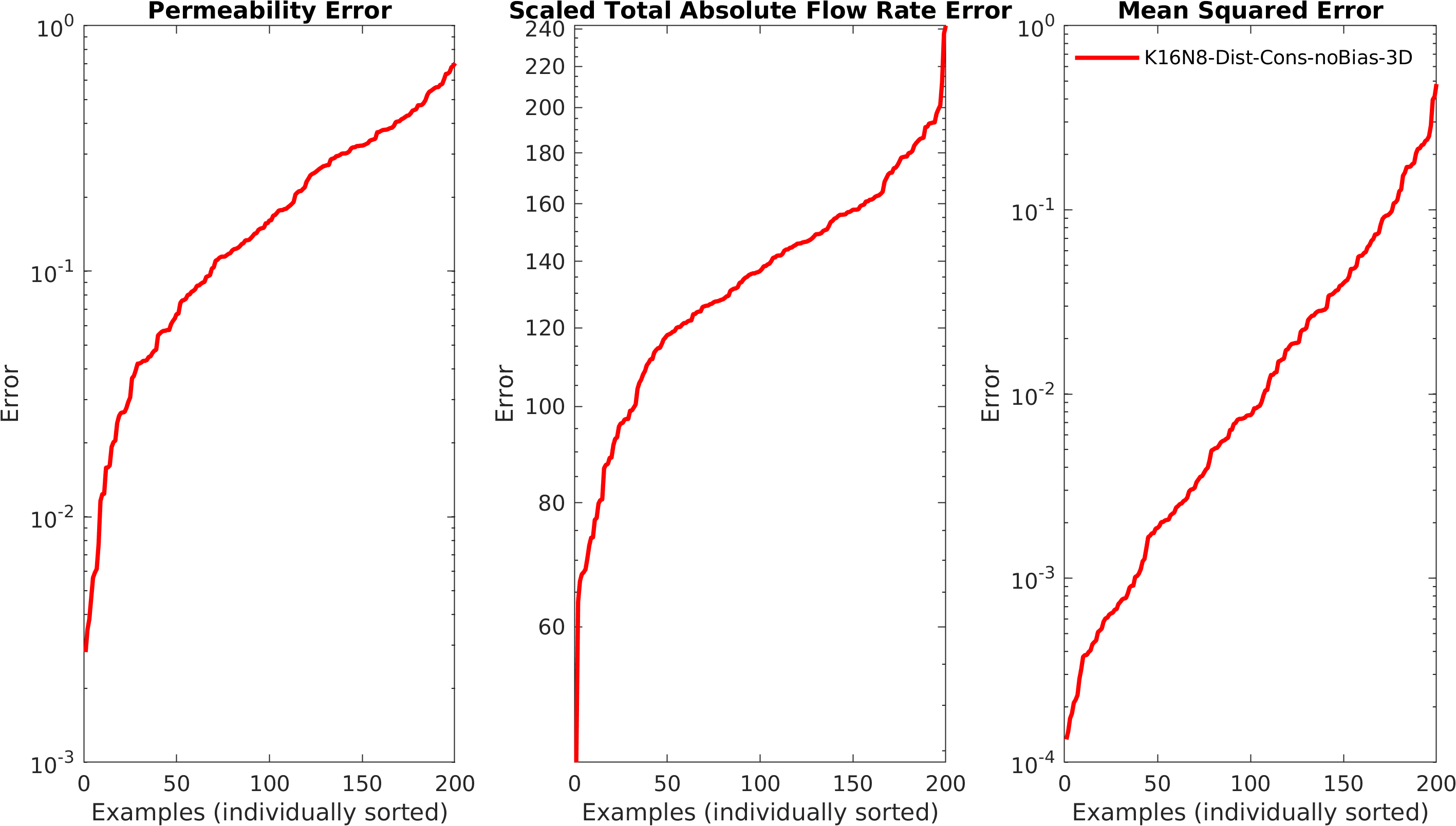}
     \begin{minipage}[b]{0.5\textwidth}
    \includegraphics[width=\textwidth]{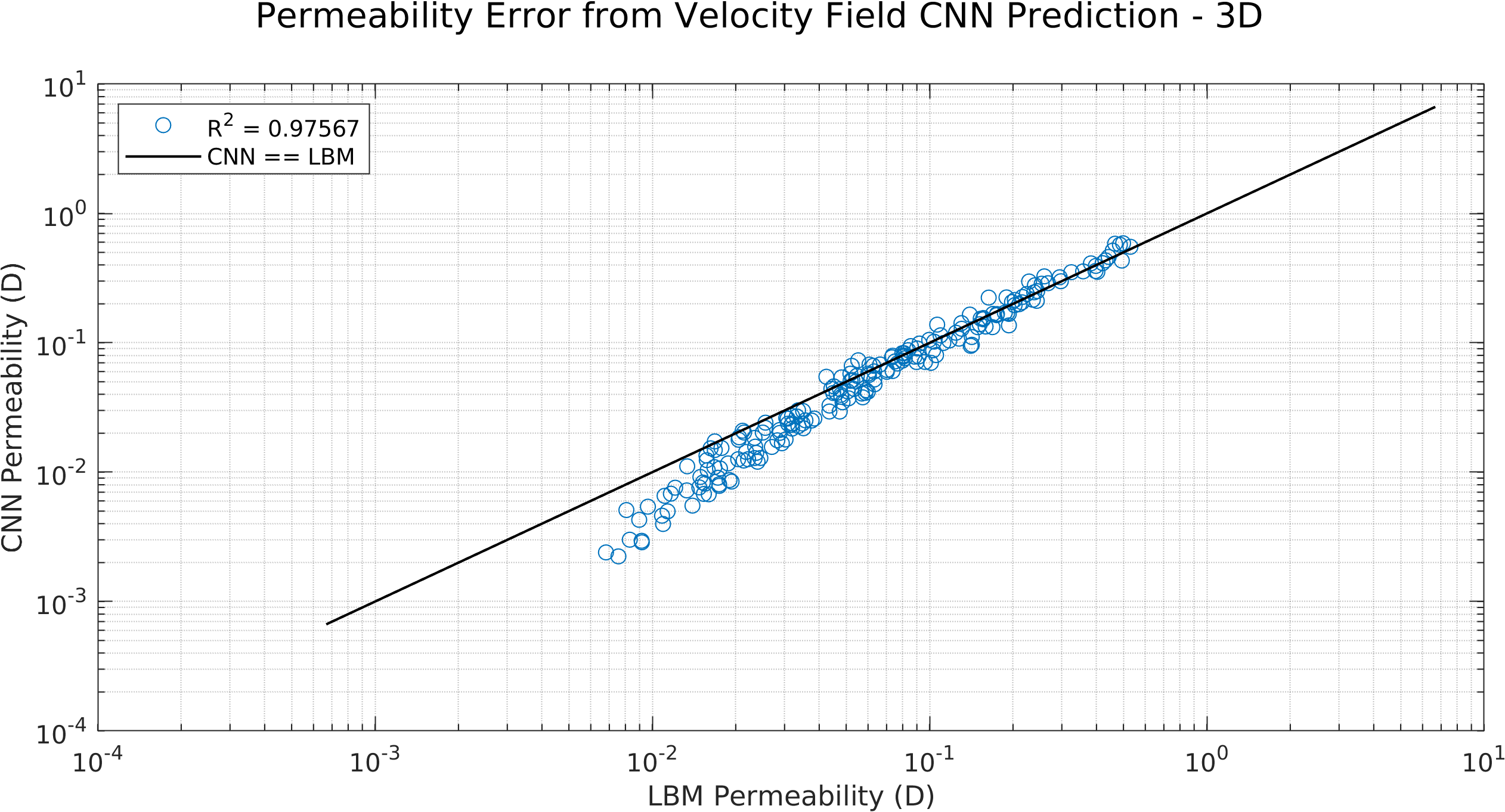}
    \end{minipage}
    \caption{Top: Plots showing the accuracy achieved by the network when predicting velocity fields, over 200 examples of 128\textsuperscript{3} correlated fields.  Due to the increase in network complexity and geometric complexity, a larger proportion of testing samples result in a permeability error higher than 10\%. Bottom: Plot of 3D testing sample permeability accuracy. Accuracy remains high for intermediate to high permeability samples, while the network underestimates values for lower permeability samples, due to the smaller values of velocity present within them.}
    \label{fig:error3dpermvperm}
\end{figure}

After having established the best configurations, the limitations in terms of accuracy, and a method of eliminating said limitations, the network is now trained and tested in 3D. Here, 1,000 correlated fields of 128\textsuperscript{3} voxels are used, 800 for training and 200 for testing. In 3D, the generated correlated fields range from simple to highly heterogeneous, and an example pair is shown in Figure \ref{fig:3dsimpcomp}. As such, much like the case in 2D where geometric complexity plays a large role in determining the accuracy of the prediction, this is more-so the case in 3D with a wider solution space. Plots shown in Figure \ref{fig:error3dpermvperm} of the permeability error, STAFE, and L2 Error show a marked decrease in achieved accuracy compared to 2D results. This is likely due to the added complexity due to the dimensionality increase, the reduction in training data, and a dataset comprised of more complex examples than in 2D. In terms of the permeability error, a majority portion of the testing samples result in predictions with an error higher than 10\%.  

A plot of the permeability predicted vs the real LBM permeability in Figure \ref{fig:error3dpermvperm} shows that accuracy is lower for samples with lower permeabilities. This is likely due to the lower overall velocity values within the domains contributing less to the training of the network. While the mass conservation loss function reduces this effect, the local minima problem inevitably creates upper limits to the accuracy achieved. These samples are 128\textsuperscript{3} correlated fields with a correlation length of around 10\% the domain length, and are thus highly tortuous domains. Examples of such domains are shown in Figure \ref{fig:3dsimpcomp}, and represent domains with a mean pore body resolved to only several voxels. This can physically occur for example in sandstone $\mu$CT images with a resolution that does not adequately resolve the pore bodies between grains. When predicting the velocity field within 3D porous media, much like the case in 2D, accuracy is limited by the geometric complexity of the domain. 

\begin{figure}[htp!]
  \centering
  \begin{minipage}[b]{0.49\textwidth}
    \includegraphics[width=\textwidth]{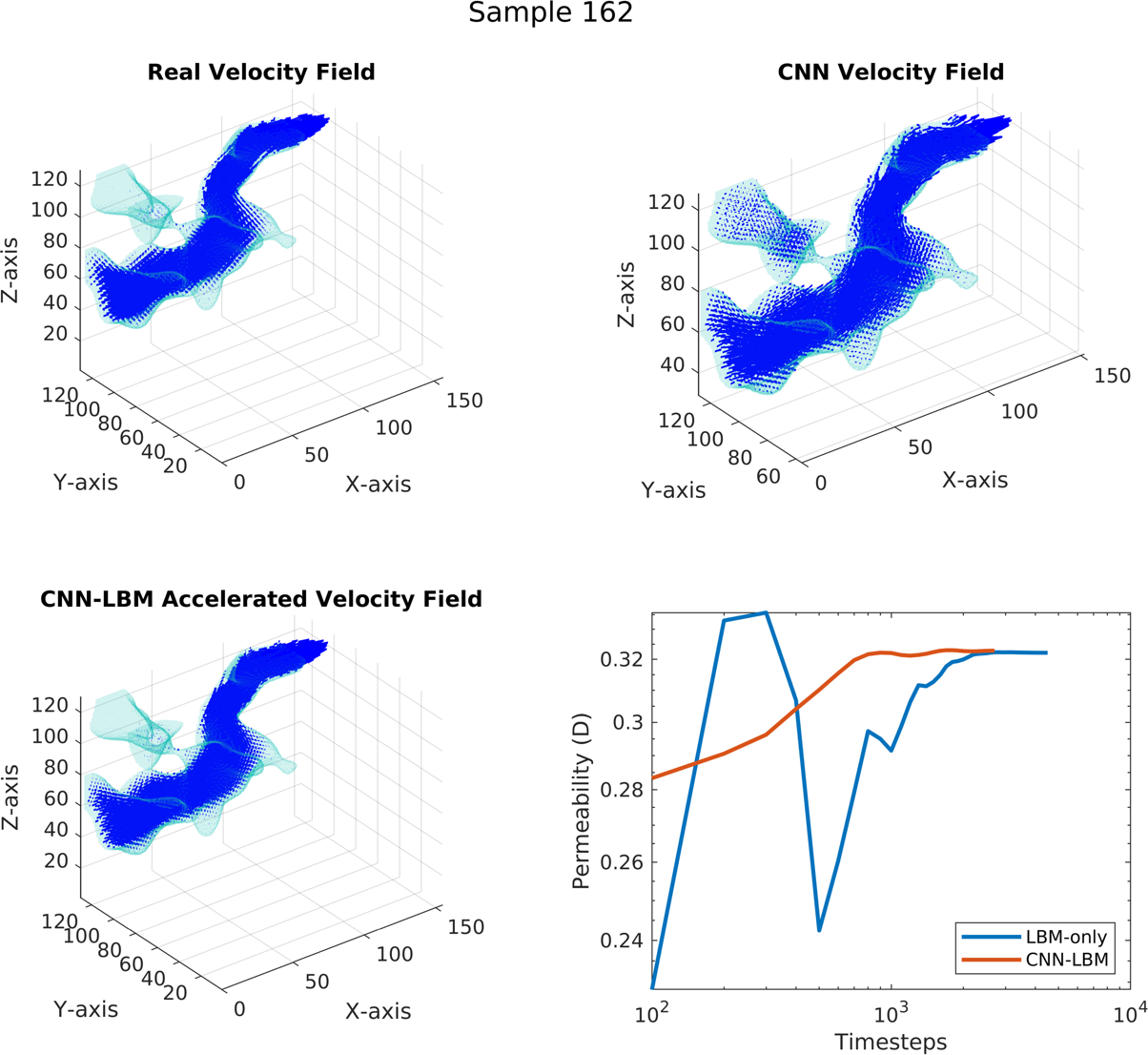}
  \end{minipage}
  \hfill
  \begin{minipage}[b]{0.49\textwidth}
    \includegraphics[width=\textwidth]{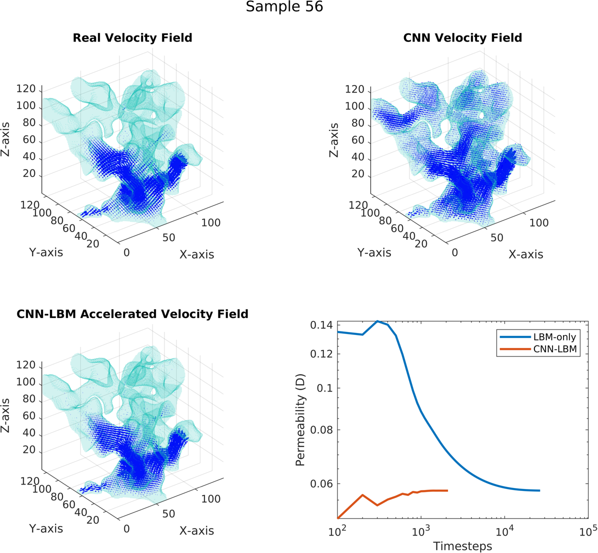}
  \end{minipage}
    \begin{minipage}[b]{0.49\textwidth}
    \includegraphics[width=\textwidth]{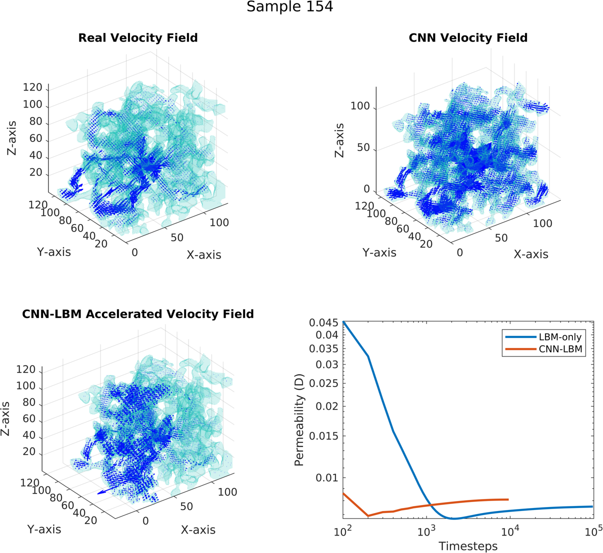}
  \end{minipage}
  \hfill
  \begin{minipage}[b]{0.49\textwidth}
    \includegraphics[width=\textwidth]{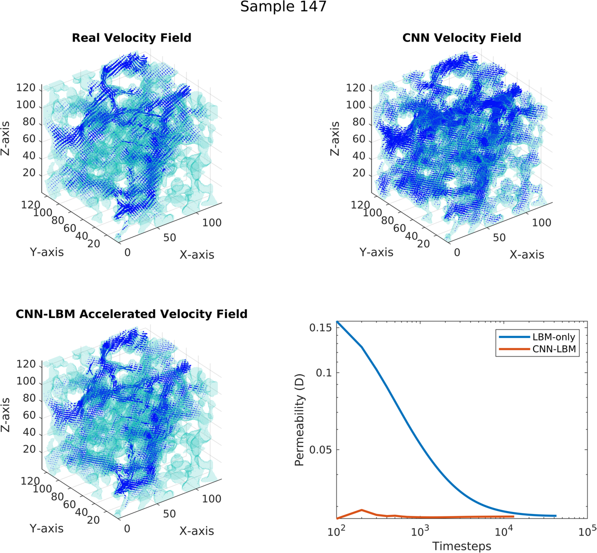}
  \end{minipage}
    \caption{Visualisation of some selected velocity fields in 3D porous media. Consistent with the results obtained in 2D testing, simple geometries result in a better match in velocity fields, while complex porous media results in predictions that struggle to ascertain the principal flow paths. Also shown are loglog plots of the acceleration achieved by preconditioning LBM with the velocity field predictions. A speedup factor of an order of magnitude is observed.}
    \label{fig:velAccelerate3d}
\end{figure}

While accuracy limitations of the network limit the use of these velocity fields as-is, accelerating direct simulation to steady state conditions using these predicted fields remains beneficial. Visualisation of the porous media with embedded velocity fields in Figure \ref{fig:velAccelerate3d} shows that errors manifest similarly to the 2D cases in previous sections. The principal flow path remains elusive when the geometry is too tortuous, and velocity values near the walls and in tight throats are inaccurate. These errors can be overcome by accelerated direct simulation, and show that the convergence rate to steady state conditions is an order of magnitude faster. 

This technique of preconditioning a direct simulation with a steady state estimate to accelerate the simulation to its true steady state conditions is a simple and generic technique. Aside from using it for correcting CNN predictions, it can potentially also be used with the outputs of Pore Network Models, Laplace Solvers, and other Navier-Stokes Estimators.



\pagebreak
\pagebreak

\section{Conclusions}
\label{sec:conclusions}

This study shows that, in the tortuous flow paths of porous media, an accurate prediction of the steady state velocity field and permeability can be obtained, with accuracy dependent on the geometric complexity. Accuracy achieved in 2D and 3D testing is consistent with previous studies in simple and porous media \cite{autodeskflow,poreflownet}. 

It is observed that the network architecture achieves a good result in permeability estimation (down to less than 1\% error in 2D and less than 10\% in 3D) by the prediction of velocity fields, which is comparable to results achieved by other velocity prediction networks and regression networks. However, the underlying velocity field is not guaranteed to possess the necessary voxel-by-voxel accuracy required for actually using these velocity fields for further analysis. Advection dominant solute transport in relatively high Peclet Numbers (near the numerical diffusion limit of Explicit Finite Volume Methods) using even the better performing predicted velocity fields in relatively simple 2D geometries shows that fine-scale errors are too large near walls and tight throats to be directly useful. 

In simpler geometries, this estimate can be used as-is for permeability estimation and garnering an approximate understanding of the flow paths. In complex geometries that the tested network design in this study struggles with, the predictions can be used as initial conditions (alongside a pressure field prediction) in direct simulation to reach a fully accurate result in a fraction of the compute time. This concept is supported by tests using LBM as the direct simulation approach, and the idea of using a "good initial guess" is also present in solving the NVE in the form of the SIMPLE (Semi-Implicit Method for Pressure-Linked Equations) algorithm for steady state flows.

Limitations resulting from the architecture of the predictor in this study are expected to be surmountable by improvements in soft computing methods of approximating velocity fields. Self correcting approximation methods are evidently the superior choice in these cases \cite{wang2019multi,hennigh2017lat}. Regardless, the efficacy of using these predictions (even the worst performing predictions), to accelerate direct simulation to steady state conditions is shown to be effective. Furthermore, the acceleration technique used in this study is general, and can be applied to any output from a flow approximator, such as Pore Network Models, Laplace Solvers, and so-on.

The use of correlated fields for the analysis of porous media is commonly used as a representation of generic geometries \cite{LIU2017121}, and acts as a proxy for the types of irregular porous media that benefit from direct simulation, such as rocks, and natural filters and membranes. The dataset is designed to be a stress test of the capabilities of CNN based predictors, and the capacity for tolerance that the acceleration method can handle while still posting reductions in compute time to steady state. 

While CNNs are focused on end-to-end mappings (raw input to final output), the use of only the distance map as an encoded input can be improved to also include other metrics as a form of Feature Engineering, such as the maximally inscribed radius \cite{poreflownet}, the Local Distance Maximum Chamber \cite{wang2019multi}, the local diffusivity (which is an output of the tortuosity calculation \cite{tortuosity}), the Time-of-Flight \cite{tof} (which is similar to the tortuosity), and so-on. These encoded inputs should be approached with caution, as some of these are surprisingly intensive to solve and scale poorly in 3D. For example, the tortuosity solved by the Algebraic Multigrid (AMG) method scales by O[N] \cite{pfvs} and the time-of-flight solved by Efficient Fast Marching scales by O[NlogN] \cite{tof}, which reduces the effectiveness of the method if a large 3D porous domain requires significant preprocessing. A middle ground between speed and input encoding should be met, and in this study, the O[N] scaling euclidean distance transform is used. It should be mentioned also that the Laplace method of approximating flow in porous media is also O[N] scaling, again emphasising that the cost of computing input parameters should be considered. 

The choice of loss functions and the use of custom loss functions can be further improved, much like the encoding of the input data. Scaling the local L2 loss or using the L1 is an option that is under-explored in this study, but has shown promise in other similar work \cite{poreflownet}. The loss in achieveable accuracy when transitioning from 2D to 3D networks is further indication that the prediction of velocity fields can be improved further. While this loss in accuracy is patched over with an acceleration routine, it would be beneficial to see further improvements in 3D results. Achieving a less than 10\% error in permeability estimation in 3D velocity field prediction provides no guarantees whatsoever that the predicted velocity fields are physically accurate or useful in further analysis that required such fine-scale accuracy. Despite these limitations, the accuracy is sufficient to be an excellent preconditioner for accelerating simulations to steady state conditions in a fraction of the otherwise compute time.

\section{Acknowledgements}
The source code used in this study is available at \url{https://github.com/yingDaWang-UNSW/VelCNNs}. 


\end{document}